\let\orienddocument\enddocument 
\documentclass[longauth]{aa}
\let\enddocument\orienddocument
\usepackage{lineno}
\usepackage{lastpage}

\usepackage{graphicx}
\usepackage[varg]{txfonts}
\usepackage{verbatim}
\usepackage{amssymb,amsmath,amsfonts}

\usepackage{threeparttable}
\usepackage{multirow} 
\usepackage{subcaption}
\usepackage{tablefootnote}

\usepackage{hyperref}
\hypersetup{
    colorlinks=true,
    linkcolor=blue,
    citecolor=blue,
    filecolor=blue,      
    urlcolor=blue,
    breaklinks=true,
}
\usepackage[normalem]{ulem} 

\usepackage{color}
\definecolor{byzantine}{rgb}{0.74, 0.2, 0.64}

\DeclareUnicodeCharacter{2212}{-}

\usepackage{natbib}
\bibpunct{(}{)}{;}{a}{}{,} 

\newcommand{\trades}[0]{\texttt{TRADES}}
\newcommand{\pyde}[0]{\texttt{PyDE}}
\newcommand{\emcee}[0]{\texttt{emcee}}
\newcommand{\pyorbit}[0]{\texttt{PyORBIT}}

\newcommand{\unif}[2]{\ensuremath{\mathcal{U} (#1,#2)}}
\newcommand{\gauss}[2]{\ensuremath{\mathcal{G}(#1,#2)}}

\begin{document}


\title{Observing a 542-day transiting giant with large TTVs}
\subtitle{The 2025 transit of HIP\,41378\,f and new constraints on the outer system}


\author{
P.~Leonardi\thanks{Corresponding author: \email{pietro.leonardi@unipd.it}}\inst{\ref{inst:unipd}, \ref{inst:inaf_oapd}}\,$^{\href{https://orcid.org/0000-0001-6026-9202}{\protect\includegraphics[height=0.19cm]{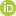}}}$
\and  A.~Santerne\inst{\ref{LAM},\ref{IPAG}}\,$^{\href{https://orcid.org/0000-0002-3586-1316}{\protect\includegraphics[height=0.19cm]{Figures/orcid.jpg}}}$
\and L.~Borsato\inst{\ref{inst:inaf_oapd}}\,$^{\href{https://orcid.org/0000-0003-0066-9268}{\protect\includegraphics[height=0.19cm]{Figures/orcid.jpg}}}$
\and S.~Grouffal\inst{\ref{LAM},\ref{IPAG}}\,$^{\href{https://orcid.org/0000-0002-2805-5869}{\protect\includegraphics[height=0.19cm]{Figures/orcid.jpg}}}$
\and C.\,R.~Mann\inst{\ref{inst:uniMo}, \ref{inst:trott}}\,$^{\href{https://orcid.org/0000-0002-9312-0073}{\protect\includegraphics[height=0.19cm]{Figures/orcid.jpg}}}$
\and G.~Piotto\inst{\ref{inst:unipd},\ref{inst:inaf_oapd}}\,$^{\href{https://orcid.org/0000-0002-9937-6387}{\protect\includegraphics[height=0.19cm]{Figures/orcid.jpg}}}$
\and K\,A.~Collins\inst{\ref{inst:cfa}}\,$^{\href{https://orcid.org/0000-0001-6588-9574}{\protect\includegraphics[height=0.19cm]{Figures/orcid.jpg}}}$
\and P.~Tamburo\inst{\ref{inst:cfa}}$^{\href{https://orcid.org/0000-0003-2171-5083}{\protect\includegraphics[height=0.19cm]{Figures/orcid.jpg}}}$ 
\and Y.~Kawai \inst{\ref{inst:tokyo}}\,$^{\href{https://orcid.org/0000-0002-0488-6297}{\protect\includegraphics[height=0.19cm]{Figures/orcid.jpg}}}$
\and D.\,C.~Stephens\inst{\ref{inst:provo}}\,$^{\href{https://orcid.org/0000-0003-4658-7567}{\protect\includegraphics[height=0.19cm]{Figures/orcid.jpg}}}$
\and J.~Garc\'ia-Mej\'ia\inst{\ref{inst:mit}, \ref{inst:cfa}}$^{\href{https://orcid.org/0000-0003-1361-985X}{\protect\includegraphics[height=0.19cm]{Figures/orcid.jpg}}}$ 
\and E.\,M.~Bryant\inst{\ref{inst:uniWarwick},\ref{inst:ceh}}\,$^{\href{https://orcid.org/0000-0001-7904-4441}{\protect\includegraphics[height=0.19cm]{Figures/orcid.jpg}}}$
\and K.\,Sz.~Zieliński\inst{\ref{inst:pol},\ref{inst:sd}}\,$^{\href{https://orcid.org/0009-0001-0389-8907}{\protect\includegraphics[height=0.19cm]{Figures/orcid.jpg}}}$
\and D.~Bayliss\inst{\ref{inst:uniWarwick},\ref{inst:ceh}}\,$^{\href{https://orcid.org/0000-0001-6023-1335}{\protect\includegraphics[height=0.19cm]{Figures/orcid.jpg}}}$
\and D.~Charbonneau\inst{\ref{inst:cfa}}\,$^{\href{https://orcid.org/0000-0002-9003-484X}{\protect\includegraphics[height=0.19cm]{Figures/orcid.jpg}}}$
\and J.\,P. ~de Leon \inst{\ref{inst:tokyo_2}}\,$^{\href{https://orcid.org/0000-0002-6424-3410}{\protect\includegraphics[height=0.19cm]{Figures/orcid.jpg}}}$ 
\and G.~Fern\'andez-Rodr\'iguez\inst{\ref{inst:tenerife},\ref{inst:ull}}\,$^{\href{https://orcid.org/0000-0003-0597-7809}{\protect\includegraphics[height=0.19cm]{Figures/orcid.jpg}}}$
\and A.~Fukui \inst{\ref{inst:tokyo_2},\ref{inst:tenerife}}\,$^{\href{https://orcid.org/0000-0002-4909-5763}{\protect\includegraphics[height=0.19cm]{Figures/orcid.jpg}}}$ 
\and E.\,G.~Hintz\inst{\ref{inst:provo}}\,$^{\href{https://orcid.org/0000-0002-9867-7938}{\protect\includegraphics[height=0.19cm]{Figures/orcid.jpg}}}$ 
\and K.~Horne\inst{\ref{inst:supa}}\,$^{\href{https://orcid.org/0000-0003-1728-0304}{\protect\includegraphics[height=0.19cm]{Figures/orcid.jpg}}}$
\and K.~Isogai \inst{\ref{inst:okayama},\ref{inst:tokyo}}\,$^{\href{https://orcid.org/0000-0002-6480-3799}{\protect\includegraphics[height=0.19cm]{Figures/orcid.jpg}}}$ 
\and J.\,S.~Jenkins\inst{\ref{inst:udp},\ref{inst:cata}}\,$^{\href{https://orcid.org/0000-0003-2733-8725}{\protect\includegraphics[height=0.19cm]{Figures/orcid.jpg}}}$  
\and N.~Narita \inst{\ref{inst:tokyo_2},\ref{inst:tokyo_3},\ref{inst:tenerife}}\,$^{\href{https://orcid.org/0000-0001-8511-2981}{\protect\includegraphics[height=0.19cm]{Figures/orcid.jpg}}}$ 
\and R.~Sefako\inst{\ref{inst:south}}\,$^{\href{https://orcid.org/0000-0003-3904-6754}{\protect\includegraphics[height=0.19cm]{Figures/orcid.jpg}}}$
\and A.~Shporer\inst{\ref{inst:mit_1}}\,$^{\href{https://orcid.org/0000-0002-1836-3120}{\protect\includegraphics[height=0.19cm]{Figures/orcid.jpg}}}$
\and R.\,~P. Schwarz\inst{\ref{inst:cfa}}\,$^{\href{https://orcid.org/0000-0001-8227-1020}{\protect\includegraphics[height=0.19cm]{Figures/orcid.jpg}}}$
\and G.~Srdoc\inst{\ref{inst:cro}}
\and P.~Wheatley\inst{\ref{inst:uniWarwick},\ref{inst:ceh}}\,$^{\href{https://orcid.org/0000-0003-1452-2240}{\protect\includegraphics[height=0.19cm]{Figures/orcid.jpg}}}$
\and J.\,B.~Williams\inst{\ref{inst:provo}}\,$^{\href{https://orcid.org/0009-0006-0339-7551}{\protect\includegraphics[height=0.19cm]{Figures/orcid.jpg}}}$
\and J.~Fern\'andez~Fern\'andez\inst{\ref{inst:uniWarwick},\ref{inst:ceh}}\,$^{\href{https://orcid.org/0000-0002-1416-2188}{\protect\includegraphics[height=0.19cm]{Figures/orcid.jpg}}}$
}

\authorrunning{Leonardi et al.}
\titlerunning{2025 Transit of HIP\,41378\,f}

    \institute{
        \label{inst:unipd}Dipartimento di Fisica e Astronomia, Università degli Studi di Padova, Vicolo dell’Osservatorio 3, 35122 Padova, Italy 
        \and 
        \label{inst:inaf_oapd}INAF, Osservatorio Astronomico di Padova, Vicolo dell'Osservatorio 5, 35122 Padova, Italy \and
        \label{LAM}Aix Marseille Université, CNRS, CNES, Institut Origines, LAM, Marseille, France \and 
        \label{IPAG}Université Grenoble Alpes, CNRS, IPAG, 38000 Grenoble, France
        \and
        \label{inst:uniMo}Département de Physique, Université de Montréal, Montréal, QC, Canada
        \and 
        \label{inst:trott}Trottier Institute for Research on Exoplanets (iREx), Montréal, QC, Canada
        \and \label{inst:cfa}Center for Astrophysics \textbar\ Harvard \& Smithsonian, 60 Garden Street, Cambridge, MA 02138, USA 
        \and \label{inst:tokyo}Department of Multi-Disciplinary Sciences, Graduate School of Arts and Sciences, The University of Tokyo, 3-8-1 Komaba, Meguro, Tokyo 153-8902, Japan
        \and \label{inst:provo}Department of Physics and Astronomy, Brigham Young University, N-486 ESC, Provo, UT 84602, USA 
        \and \label{inst:mit}~Kavli Institute for Astrophysics and Space Research, Massachusetts Institute of Technology, Cambridge, MA 02139, USA 
        \and \label{inst:uniWarwick}Department of Physics, University of Warwick, Gibbet Hill Road, Coventry CV4 7AL, United Kingdom 
        \and\label{inst:ceh}Centre for Exoplanets and Habitability, University of Warwick, Coventry, CV4 7AL, UK
        \and \label{inst:pol}Obserwatorium Astronomiczne Niedźwiady, Szubin, Poland
        \and \label{inst:sd}Boyce Research Initiatives and Education Foundation, San Diego, CA, USA
        \and \label{inst:tokyo_2}Komaba Institute for Science, The University of Tokyo, 3-8-1 Komaba, Meguro, Tokyo 153-8902, Japan 
        \and \label{inst:tenerife}Instituto de Astrof\'{i}sica de Canarias (IAC), 38205 La Laguna, Tenerife, Spain
        \and \label{inst:ull}Departamento de Astrof\'isica, Universidad de La Laguna (ULL), 38206 La Laguna, Tenerife, Spain 
        \and \label{inst:supa}SUPA Physics and Astronomy, University of St. Andrews, Fife, KY16 9SS Scotland, UK
        \and \label{inst:okayama}Okayama Observatory, Kyoto University, 3037-5 Honjo, Kamogatacho, Asakuchi, Okayama 719-0232, Japan
        \and \label{inst:udp}Instituto de Estudios Astrof\'isicos, Facultad de Ingenier\'ia y Ciencias, Universidad Diego Portales, Av. Ej\'ercito 441, Santiago, Chile 
        \and \label{inst:cata}Centro de Astrof\'isica y Tecnolog\'ias Afines (CATA), Casilla 36-D, Santiago, Chile 
        \and \label{inst:tokyo_3}Astrobiology Center, 2-21-1 Osawa, Mitaka, Tokyo 181-8588, Japan
        \and \label{inst:south}South African Astronomical Observatory, P.O. Box 9, Observatory, Cape Town 7935, South Africa
        \and \label{inst:mit_1}Department of Physics and Kavli Institute for Astrophysics and Space Research, Massachusetts Institute of Technology, Cambridge, MA 02139, USA
        \and  \label{inst:cro}Kotizarovci Observatory, Sarsoni 90, 51216 Viskovo, Croatia
}

\date{Received 8 April 2026 / Accepted  26 June 2026}

\abstract{
Characterizing long-period transiting exoplanets is inherently challenging due to the rarity and long duration of transit events. Yet, these systems provide unique insights into planetary formation, migration, the detection of exomoons, and primordial atmospheres by occupying a sparsely populated region of the exoplanet parameter space. The complexity increases further for long-period planets near mean-motion resonances, where transit timing variations (TTVs) can reach amplitudes of several hours to days. We present a coordinated space- and ground-based observing campaign, using photometry from NEOSSat, multiple LCOGT sites, MuSCAT, MuSCAT3, Tierras, and NGTS, to capture the 19-hour transit of the long-period giant exoplanet HIP\,41378\,f ($P \approx$  542 d, $R \approx$ = 9.5 $R_{\oplus}$) on 31 October 2025. Our transit analysis constrains the time of inferior conjunction to $T_{\mathrm{C}} = 2460980.888 \pm 0.029~\mathrm{BJD_{TDB}}$, occurring $\sim 7$ hours earlier than predicted from its linear ephemeris. 
This significant offset is consistent with the previously reported TTVs of HIP\,41378\,f, making it the longest-period exoplanet known to exhibit measurable TTVs.
By combining this new precise measurement to the transit timings of the two outer planets in the system (HIP\,41378\,d and HIP\,41378\,e), we perform a dynamical modeling of the system, using the N-body integrator \trades, refine the ephemeris of HIP\,41378\,f, and predict future transit events for all three outer transiting planets.
}


\maketitle

\nolinenumbers

\section{Introduction}
\label{section:intro}
Even with more than 6\,000 exoplanets discovered to date, the current census of planetary systems remains substantially incomplete. This is largely due to observational biases inherent to the most widely used detection techniques, namely transit photometry and radial-velocity spectroscopy, which strongly favor short-period planets with frequent transits or large radial-velocity amplitudes \citep{winn_2015}. This bias is particularly severe for transiting planets, where the geometric transit probability decreases with orbital separation and the detection efficiency is further limited by the finite temporal baselines of space-based surveys such as \textit{Kepler} and TESS, as well as their long-transit duration. As a consequence, long-period, transiting temperate giant planets ($P \gtrsim 300$\,d; $T_\mathrm{eq} \lesssim 400$ K) are rare in the current exoplanet sample. 
To date, among the more than $\sim$4600 confirmed planetary systems listed in the NASA Exoplanet Archive\footnote{\url{https://exoplanetarchive.ipac.caltech.edu/} accessed on 26 May 2026}, only a small fraction host long-period transiting temperate giant planets with measured radii and orbital periods. Among these, only six high-multiplicity systems ($N_{\mathrm{planets}} > 3$) are currently known: HIP\,41378 \citep{Grouffal_2026}, KOI-351 \citep{cabrera_2014}, Kepler-62 \citep{borucki_2013}, Kepler-89 \citep{weiss_2013}, Kepler-150 \citep{rowe_2014}, and Kepler-167 \citep{rowe_2014, kipping_2016}.

Despite their rarity, long-period planets are of particular interest because they probe dynamical and physical regimes that remain largely unexplored. For instance, planets with orbital periods longer than $\sim$300 days have a high probability ($\gtrsim 60\%$) of retaining primordial satellites on long-term stable orbits within their Hill spheres \citep{gong_2013, Dobos_2021}, making the monitoring of their transits fundamental for detecting potential massive exomoons. At the same time, the atmospheres of long-period planets offer a unique opportunity to probe chemical and dynamical regimes that are largely inaccessible for close-in systems. Their cooler equilibrium temperatures allow primordial atmospheres to persist largely unaffected by photoevaporation and core-powered mass loss, thereby preserving information about their formation and early evolution \citep{owen_Wu_2017, ginzburg_2018, Fortney_2020}.

In this scientific landscape, the planetary system orbiting BD+10\,1799 (HIP\,41378) represents a particularly valuable benchmark. The host star is an F7V star ($T_\mathrm{eff} = 6290 \pm 77$\,K), originally cataloged in the 19th century by Friedrich Wilhelm Argelander in the \textit{Bonner Durchmusterung des nördlichen Himmels} under the designation BD+10\,1799 \citep{Argelander_1903}, before later receiving its \textit{Hipparcos} identifier \citep{Perryman_1997} (See Table~\ref{table:stellar_parameters}). Hosting multiple transiting planets with long orbital periods, the system enables a rare joint investigation of planetary atmospheres, bulk compositions, and dynamical architecture within a single system \citep{vanderburg_2016, santerne_2019, alam_2022, Leonardi_2025, Grouffal_2026}.

The planetary system comprises five transiting planets and one non-transiting companion, with orbital periods spanning from 15 to 542 d \citep{santerne_2019}. According to the results of \cite{Leonardi_2025} and \cite{Grouffal_2026}, the system seems to be divided into two dynamically decoupled sub-systems. The inner sub-system consists of a compact triplet of planets near a 4:2:1 zero-order three-body resonant chain, with orbital periods of $\rm P_b = 15.57$ d, $P_\mathrm{c} = 31.71$ d, and $P_\mathrm{g} = 64.07$\,d \citep{Leonardi_2025}. In contrast, the dynamical architecture of the outer sub-system remains less constrained. Current observations are consistent with a configuration in which the three outer giant planets occupy a second-order near-resonant chain; specifically, the adjacent pairs d–e and e–f lie close to second-order 7:5 two-body mean-motion resonances.
The orbital periods of $P_\mathrm{d} \sim 278$\,d, $P_\mathrm{e} \sim 386$–390\,d, and $P_\mathrm{f} = 542$\,d, as suggested by \cite{Grouffal_2026}, correspond to period ratios of $P_{\rm e}/P_{\rm d} \sim 1.396$ and $P_{\rm f}/P_{\rm e} \sim 1.397$, both sitting within 0.3\% of the exact 7:5 commensurability ($=1.40$)

Thanks to more than a decade of radial-velocity monitoring, the HIP\,41378 system has been extensively characterized, yielding robust mass constraints for planets b, c, g, and f, while also revealing evidence for a putative outer companion, namely HIP\,41378\,h \citep{santerne_2019, Grouffal_2026}.
However, both planets d and e suffer from poorly determined ephemeris and orbital periods, due to the limited number of observed transits. In addition, the available radial-velocity data do not yet provide a well-constrained Keplerian reflex signal, and their masses therefore remain poorly determined \citep{Grouffal_2026}.

In this work, we present the photometric observation of the 2025 transit of HIP\,41378\,f ($ P_\mathrm{f}$ = 542.08\,d, $R_\mathrm{f}$ = 9.4 $R_{\oplus}$, $M_\mathrm{f}$ = 25 $M_{\oplus}$) obtained through a coordinated campaign combining space-based and ground-based facilities. Owing to its long orbital period, HIP\,41378\,f exhibits an exceptionally long transit duration of approximately 19 hours \citep{santerne_2019, Grouffal_2026}. As a result, capturing a complete transit from a single ground-based observatory is impractical, necessitating either a global, multi-site observing strategy across different longitudes or continuous space-based observations, or a combination of both.
The observation of the transit of this planet is further complicated by the presence of significant TTVs, which introduce substantial uncertainty in the predicted transit times \citep{bryant_2021, alam_2022, Garcia_Meija_2025}. These TTVs increase the risk of partial transit coverage or entirely missed transit events and, therefore, require both flexible scheduling and precise coordination among observing facilities. To overcome this issue and help planning follow-up campaigns of the three outer transiting planets in the system (HIP\,41378\,d, HIP\,41378\,e, and HIP\,41378\,f) we dynamically model their TTVs using a N-body integrator, providing refined constraints on their orbital periods and predicted transit windows.

The structure of this paper is as follows. In Sect.~\ref{sec:obs}, we present the new and archival photometric observations used in our analysis. The methodologies for our photometric and dynamical modeling are detailed in Sects.~\ref{sec:photo_fit} and \ref{sec:dyna_fit}, respectively. We provide our results and a broader dynamical discussion in Sect.~\ref{sec:resu}, before concluding with a summary of our findings in Sect.~\ref{sec:conclusions}.

\begin{table}[tb]
\centering\centering\renewcommand{\arraystretch}{1.3}\small
    \caption{Stellar properties of HIP\,41378 (BD+10 1799).} 
    \begin{tabular}{l c c }
    \hline\hline
    Parameter & Value& References \\
    \hline
    RA [J2000] & 08 26 27.85 &  2\\
    Dec [J2000] & +10 04 49.3 &  2\\
    $B$ [mag] & $ 9.42 \pm 0.20$ & 3 \\
    $V$ [mag] & $8.92  \pm 0.20$ & 3 \\
    $G$ [mag] & $8.82 \pm 0.0028$ & 2 \\
    $J$ [mag] & $7.98 \pm 0.026$ & 4 \\
    $H$ [mag] & $7.79 \pm 0.038$ & 3 \\
    $K$ [mag] & $7.72 \pm 0.031$ & 3 \\
    \\
    
    T$_\mathrm{eff}$ [$K_s$] & $ 6290 \pm 77 $& 1 \\
    $\log g$ [cgs] & $4.298 \pm 0.004 $ & 1\\
    $[\mathrm{Fe}/\mathrm{H}]$\,[dex] & $ -0.05\pm 0.10$ & 1\\
    $M_{\star}$ [$M_{\odot}$] & $1.22^{+0.03}_{-0.02}$ & 1\\ 
    $R_{\star}$ [$R_{\odot}$] & $1.300 \pm 0.009 $ & 1\\
    $\rho_{\star}$ [ g cm$^{-3}$] & $0.785 \pm 0.008$ & 1\\
    Age [Gyr] & $2.07_{-0.27}^{+0.36}$ & 1\\
    d (pc) & $106.8 \pm 1.0$ & 1 \\

    \hline
    \end{tabular}
    \tablebib{(1) \cite{Lund_2019}. (2) \textit{Gaia} DR3 \citep{gaia_DR3}. (3) Tycho-2 Catalogue \citep{tycho-2}. (4) Two Micron All Sky Survey \citep[2MASS][]{skrutskie_2006}.}
    \label{table:stellar_parameters}
\end{table}
 
\section{Observations}
\label{sec:obs}

\subsection{2025 Transit Campaign}
We observed a transit of HIP\,41378\,f between 31 October 2025 and 1 November using an ensemble of space- and ground-based facilities, including NEOSSat, multiple LCOGT sites, MuSCAT, MuSCAT3, the BYU 24-inch telescope, Tierras, and NGTS.
The observing campaign was scheduled based on the transit time predicted by \cite{Garcia_Meija_2025}, $T_{\mathrm{C,pred}} = 2460980.793^{+0.098}_{-0.129}~\mathrm{BJD_{TDB}}$. The ground-based observations were largely coordinated within the \textit{TESS} Follow-up Observing Program Sub Group 1 \citep[TFOP SG1;][]{collins:2019}\footnote{\url{https://tess.mit.edu/followup}}.

Thanks to the NEOSSat long-observational window, we were able to detect the full transit event (see Fig.~\ref{fig:lc_neossat}). However, its photometry is affected by significant data gaps-due to the periodic Earth occultations of the spacecraft-as well as a relatively large photometric scatter, preventing continuous coverage of the ingress and egress phases. By combining these observations with data from a global network of ground-based observatories, we significantly improved the temporal coverage of the transit, particularly during the egress, thereby enhancing the precision on the inferred time of inferior conjunction.

\paragraph{NEOSSat.} 
The Near-Earth Object Surveillance Satellite (NEOSSat; \citealt{Laurin_2008}) is a Canadian microsatellite equipped with a 15-cm aperture Maksutov telescope feeding a back-illuminated CCD detector, providing a field of view of approximately 51.6\,arcmin~$\times$~51.6\,arcmin with a plate scale of $\sim$3.0\,arcsec per pixel. NEOSSat observed a full transit of HIP\,41378\,f during a $\sim$3-day window based on the predicted time of inferior conjunction ($T_{\rm{{C, pred}}}$), spanning 31 October 2025 (10:10 UTC) to 3 November 2025 (12:04 UTC).
The observations were obtained without a photometric filter. The NEOSSat raw images were reduced using a dedicated Python-based pipeline\footnote{\url{https://github.com/jasonfrowe/neossat}}, following the procedure described in \citet{Mann_2023}. Aperture photometry was extracted for the target and a set of comparison stars in the field. A principal component analysis was then applied to the raw photometry using 20 reference stars to construct a normalized relative flux light curve for the target. The principal component analysis procedure removes time-dependent instrumental and environmental trends that are common among stars in the field. In our analysis, the first five principal components were used to model and subtract these shared systematics. Residual low-frequency variability not captured by the principal component analysis was subsequently removed using spline detrending, resulting in a final light curve that clearly reveals the transit signal within the predicted transit window.

\paragraph{LCOGT.} We observed time windows of HIP\,41378 (TOI-4304) on 1 November 2025 November from the Las Cumbres Observatory Global Telescope \citep[LCOGT;][]{Brown:2013} 0.35\,m and 1.0\,m network nodes at Teide Observatory on the island of Tenerife (Teide), McDonald Observatory near Fort Davis, Texas, United States (McD), Haleakala Observatory on Maui, Hawai'i (Hal), and Siding Spring Observatory near Coonabarabran, Australia (SSO). We observed in Sloan $i'$ band with the 0.35\,m telescopes and Pan-STARRS $z_s$ band with the 1.0\,m telescopes. The 0.35\,m Planewave Delta Rho 350 telescopes are equipped with a $9576\times6388$ QHY600 CMOS camera having an image scale of $0.73$\,arcsec per pixel, resulting in a 114\,arcmin~$\times$~72\,arcmin full field of view. We used the optional 30\,arcmin~$\times$~30\,arcmin sub field of view for a faster detector read-out. The 1\,m telescopes are equipped with $4096\times4096$ SINISTRO cameras having an image scale of $0\farcs389$ per pixel, resulting in a 26\,arcmin~$\times$~26\,arcmin field of view. The images were calibrated by the standard LCOGT {\tt BANZAI} pipeline \citep{McCully:2018}, and photometric data were extracted using {\tt AstroImageJ} \citep{Collins:2017}. Circular apertures with 6--7\,arcsec and 8--11\,arcsec radii were used to extract the differential photometry for the 1\,m and 0.35\,m data, respectively. All data were determined to be in-transit, with the start of egress occurring during the SSO observation window (see Section \ref{sec:photo_fit}).

\paragraph{BYU 24-inch.} HIP\,41378 was observed on 1 November 2025 with an SBIG STX-16803 CCD camera, mounted on a 0.6~m Planewave CDK24 telescope, at the Orson Pratt Observatory in Provo, Utah, USA.  The CCD has $4096 \times 4096$ pixels, a plate scale of $0.47$\,arcsec per pixel, and a field of view of 32\,arcmin~$\times$~32\,arcmin. Observations were taken in the $I$ filter, initially employing an integration time of 60\,s, which was subsequently decreased to 45\,s to optimize the signal-to-noise ratio while strictly avoiding detector saturation. The observations were carried out from 09:19 UT until 12:56 UT. The first observation occurred at an airmass of 2.00, and the last observation was taken at an airmass of 1.16.  The raw data were reduced using an in-house \texttt{astropy} script that processes and applies the bias, dark, and flat-field calibration frames to the data. 

\paragraph{MuSCAT \& MuSCAT3.}
HIP\,41378 was observed using the Multi-color Simultaneous Camera for studying Atmospheres of Transiting exoplanets (MuSCAT) series \citep{narita_2015, narita_2020} on 1 November 2025. The MuSCAT and MuSCAT3 instruments are part of the series of multi-color, simultaneous-imaging photometers designed for high-precision transit observations, installed on the 1.88 m telescope at the National Astronomical Observatory of Japan in Okayama, Japan and the and the 2\,m Faulkes Telescope North of LCOGT on Haleakalā, Maui, respectively. 

MuSCAT is equipped with three 1024 $\times$ 1024 pixel CCDs intended for each of the $g$ (400–550 nm), $i$ (700–820 nm) and $z_s$ (820–920 nm) observing bands, with a field of view of 6.1\,arcmin~$\times$~6.1\,arcmin and pixel scale of 0.358\,arcsec per pixel. MuSCAT3 is hosts four 2048 $\times$ 2048 pixel CCDs, with the addition of $r$ (550–700\,nm) band, as well as the choice to opt for narrow-band filters \citep{fukui_2024}, which are $g_\mathrm{narrow}$ (490–530 nm), 
\textit{$Na_D$} (583--595 nm; centered at the sodium absorption line),
$i_\mathrm{narrow}$ (776–808\,nm) and $z_\mathrm{narrow}$ (846–890\,nm). The field of view for MuSCAT3 is 9.1\,$\times$\,9.1\,arcmin and the pixel scale is 0.27\,arcsec\, pixel$^{-1}$,

With MuSCAT, we observed with the exposure time of 5s in all bands with a heavy defocus to avoid saturation. The observation was conducted from 15:40 UT, but weather limited the observation to the first 50 minutes. With MuSCAT3, we deployed the narrow-band filters on top of heavy defocus to avoid saturation. The exposure time was 6\,s, 5\,s, 7\,s, 16s for the respective bands, and the observation was conducted from 12:30 UT to 15:30 UT with clear weather throughout.

For MuSCAT, the data were manually reduced by subtracting a master dark and dividing by master flat frame. For MuSCAT3, the data was automatically reduced by the LCOGT \texttt{BANZAI} pipeline. We performed aperture photometry on all datasets using a custom  pipeline that optimizes both the aperture radius and the choice of comparison stars. We iteratively discard $3\sigma$ outliers from a second-order polynomial fit in the raw flux of the brightest comparison star in this process. The optimization was then based on a merit function that rewards lower point-to-point scatter in the resulting light curves while penalizing an increased fraction of discarded data points. Further detrending was also performed by fitting the observed flux as a linear function of the centroid position on the detector ($\Delta x$ and $\Delta y$) for both light curves.

\paragraph{Tierras.} 
HIP\,41378 was observed with Tierras on four consecutive nights from 30 October 2025 to 2 November 2025. The Tierras Observatory is a robotic 1.3\,m telescope located atop Mt. Hopkins at Fred Lawrence Whipple Observatory in southern Arizona. It is equipped with an ultra-precise photometer with a custom narrow-band NIR filter ($\lambda=863.5$\,nm, $\text{FWHM}=40.2$\,nm; \citealt{Garcia_Mejia_2020}). Observations were taken with an 8\,s exposure time over the entire hour-angle range where the target was observable, resulting in approximately four hours of coverage on each night. 

The data were processed with the Tierras photometric pipeline \citep{Tamburo_2025}, which performs the data calibration and light curve creation. The final light curve used a circular aperture with a radius of 10\,pixels. We applied a quality mask to this light curve, removing any exposures with World Coordinate System solutions with an root-mean-square greater than 0.215\,arcsec (half a Tierras pixel), $x$ or $y$ pointing deviations greater than 20~pixels, FWHM seeing values greater than 4\,arcsec, and average reference star flux values less than 90\% of the median value (to eliminate exposures with significant cloud cover). We also discarded any exposures where the photometric aperture centered on HIP\,41378 contained pixels with more than $40\,000$~counts, to avoid non-linearity effects, and performed a $5\sigma$ clipping to remove any remaining extreme outliers. 

We did not catch the ingress or egress of HIP~41378~f in the Tierras light curve. However, upon inspection, we observed that the data taken on 1 November 2025 were decremented with respect to the other three nights by an amount that is consistent with the expected transit depth of HIP\,41378\,f. Suggesting that the star was observed while the planet was mid-transit during the third night of the Tierras campaign, a scenario that occurred during a previous Tierras observation of a transit of HIP\,41378\,f \cite{Garcia_Meija_2025}.  We therefore re-normalized the Tierras light curve by the median flux value of the first, second, and fourth nights in the dataset, when the planet was apparently out of transit. 

\paragraph{NGTS.}
We monitored HIP\,41378 over seven consecutive nights, from 29 October 2025 to 4 November 2025 inclusive. Observations were conducted using a consistent setup of six NGTS telescopes throughout the entire duration of the campaign.
The Next Generation Transit Survey \citep[NGTS; ][]{Wheatley_2018_ngts} is a ground-based photometric facility situated at the European Southern Observatory Paranal Observatory in Chile and specialized for the high-precision observation of exoplanet host stars. The NGTS facility consists of twelve independently steerable telescopes, each with a 20\,cm diameter aperture, a 156\,arcmin~$\times$~156\,arcmin field of view field-of-view, and a deep-depleted red sensitive CCD \citep{Wheatley_2018_ngts}. Using multiple NGTS telescopes to simultaneously observe the star has been shown to yield a significant increase in the photometric precision achieved by NGTS \citep{Smith_2020_multicam, Bryant_2020_multicam}. This combination of high photometric precision from multiple cameras and the wide field-of-view of each individual camera providing a large selection of comparison stars with which to correct systematic trends enables NGTS to achieve sub-part-per-thousands level precision photometry for very bright host stars from the ground \citep{bryant_2021, Yu_2025}. For example, for the 2019 11-camera observations of the transit HIP-41378f reported in \citet{bryant_2021} the NGTS observations achieved a photometric precision of 162\,ppm on a timescale of 30\,minutes. The NGTS cameras also utilize the \texttt{DONUTS} guiding algorithm \citep{McCormac_2013} to achieve sub-pixel level stability on the telescope guiding across timescales of months, which in turn provides excellent photometric night-to-night stability.

The photometric image reduction was performed using a custom aperture photometry pipeline that uses the \texttt{SEP} software for source extraction and photometry, where \texttt{SEP} is a Python implementation of the \texttt{SExtractor} algorithm \citep{Bertin_1996, Barbary_2016}. Following the procedure described by \citet{Bryant_2020_multicam} the \textit{Gaia} catalog information \citep{gaia_DR2, GaiaCollaboration2023} was used to select the set of unblended comparison stars closest to the target in terms of apparent magnitude, position on the CCD, and stellar color. The set of comparison stars used was kept the same for all observation nights in order to ensure that we did not introduce any spurious night-to-night photometric offsets during this step. The set of comparison stars was then used to produce the differential flux time series for each camera.

\begin{table*}
\centering
\small
\renewcommand{\arraystretch}{1.2}
\setlength{\tabcolsep}{4pt}
\caption{Transit fit derived planetary and orbital parameters for HIP\,41378\,d, e, and f.}
\begin{tabular}{l cc cc cc}
\hline\hline
 & \multicolumn{2}{c}{HIP\,41378\,d} 
 & \multicolumn{2}{c}{HIP\,41378\,e} 
 & \multicolumn{2}{c}{HIP\,41378\,f} \\
Parameter 
& Prior & Value 
& Prior & Value 
& Prior & Value \\
\hline

$ P$ [d] 
& \gauss{278.3618}{0.0003}
& $278.36180_{-0.0003}^{+0.0003}$ 
& \unif{360}{415} 
& $388.4_{-7.6}^{+11.0}$ 
& \gauss{542.0797}{0.0001}
& $542.07970_{-0.0001}^{+0.0001}$ \\

$ b$ 
& \gauss{0.54}{0.03} 
& $0.548_{-0.028}^{+0.024}$ 
& \gauss{0.632}{0.039}
& $0.589_{-0.035}^{+0.035}$ 
& \gauss{0.207}{0.034}
& $0.197_{-0.028}^{+0.027}$ \\

$ R_p/R_\star$ 
& \gauss{0.02573}{0.00050}
& $0.02563_{-0.00036}^{+0.00034}$ 
& \gauss{0.03648}{0.00014} 
& $0.03655_{-0.00014}^{+0.00014}$ 
& \gauss{0.06656}{0.00011} 
& $0.066603_{-0.000098}^{+0.000099}$ \\

$ R_{p}$ [$R_{\oplus}$] 
& ...
& $3.635_{-0.057}^{+0.055}$
& ...
& $5.184_{-0.041}^{+0.041}$
& ...
& $ 9.448_{-0.067}^{+0.067}$\\

$ i$ [deg]
& \gauss{89.789}{0.005}
& $89.7831_{-0.0085}^{+0.0078}$
& \gauss{89.801}{0.003}
& $89.8062_{-0.0072}^{+0.0067}$
& \gauss{89.948}{0.008}
& $89.9507_{-0.0066}^{+0.0070}$\\

$ e$
& \unif{0.0}{0.5}
& $<0.24$
& \unif{0.0}{0.5}
& $<0.28$
& \unif{0.0}{0.5}
& $<0.15$\\

$ T_{14}$ [d]
& ...
& $0.5210_{-0.0094}^{+0.011}$
& ...
& $0.573_{-0.017}^{+0.016}$
& ...
& $0.7874_{-0.0068}^{+0.0056}$\\

\hline
\end{tabular}
\tablefoot{For each parameter, the adopted prior distribution and posterior value are reported. Uniform ($\mathcal{U}$) and Gaussian ($\mathcal{G}$) priors are used as indicated.}
\label{tab:Transit_fit_results}
\end{table*}

\begin{table}[]
    \small\centering\renewcommand{\arraystretch}{1.3}
    \addtolength{\tabcolsep}{-0.45em}
    \caption{Transit times of HIP\,41378\,d, HIP\,41378\,e, and HIP\,41378\,f.}
    \label{tab:T0s}
    
    \begin{threeparttable} 
    \begin{tabular}{l c c c}
    \hline\hline
    $ T_0$ ($\mathrm{BJD_{TDB}}$) & $\sigma_{T0}$ (d) &  Telescope  & Source\\
    \hline
    \emph{HIP\,41378\,d} \rule{0pt}{12pt} & & & \\
    $2\,457\,166.2698$ & $0.0018$ & K2 & This work \\ 
    $2\,458\,279.7070$ & $0.0021$ & K2 & This work \\

    \emph{HIP\,41378\,e} \rule{0pt}{12pt} & & & \\
    $2\,457\,142.01570$ & $0.00064$ & K2 & This work \\

    \emph{HIP\,41378\,f} \rule{0pt}{12pt} & & & \\
    $2\,457\,186.91477$ & $0.00053$ & K2 & This work \\
    $2\,458\,271.07523$ &  $0.00077$ &  K2 & This work \\
    $2\,458\,813.0913$ & $0.0046$ & NGTS & \cite{bryant_2021} \\
    $2\,459\,355.10066$ &  $0.00054$ & HST & This work \\
    $2\,459\,897.0196$ & $0.0009$ & Spectrographs & \cite{Grouffal_2025} \\
    $2\,460\,438.891$ &   $0.052$\tnote{a} &  Tierras & \cite{Garcia_Meija_2025} \\
    $2\,460\,980.888$  & $0.029$  &  Global & This work \\
    \hline
    \end{tabular}
    
    \begin{tablenotes} 
        \small
        \item[a] This timing corresponds to the median value of a uniform posterior distribution.
    \end{tablenotes}
    \end{threeparttable}
\end{table}

\subsection{Literature observations}

\paragraph{K2.} 
The target was observed during Campaigns~5 and~18 of the K2 mission \citep{howel_2014}. Across these two campaigns, two transits of HIP\,41378\,f, one of HIP\,41378\,e, and two of HIP\,41378\,d were captured. We retrieved the high-level science data products from the Mikulski Archive for Space Telescopes\footnote{\url{https://mast.stsci.edu/portal/Mashup/Clients/Mast/Portal.html}}, which were generated using the \texttt{EVEREST} photometric pipeline \citep{luger_2016,luger_2018}. From these light curves, we extracted segments centered on each transit.

\paragraph{HST.}
We used the public available Hubble Space Telescope (HST) WFC3 \citep{marinelli_2024} transit observations of HIP\,41378\,f from the MAST archive. 
The dataset consists of a single transit observed with the G141 grism (1.125–1.643~$\mu$m) as part of program GO~16267 (PI: Dressing), acquired between 19 and 21 May 2021 over three HST visits. These observations were previously published by \citet{alam_2022}. Each visit comprised six HST orbits, with individual exposures of 37\,s obtained at a scan rate of 0.25\,arcsec and using eight up-the-ramp samples per exposure.
For wavelength calibration, a direct image of the target was taken during each visit using the \textit{F126N} filter. Following standard practice, we excluded the first orbit of every visit from the analysis due to the presence of strong time-dependent systematics. The observations were performed with the GRISM256 aperture and the SPARS10 readout sequence, employing both forward and reverse spatial scanning directions. The raw WFC3 data were calibrated and the white light curves extracted using the dedicated \texttt{Iraclis} pipeline \citep{Tsiaras_2016a, Tsiaras_2016b, Tsiaras_2018}. Light-curve extraction was carried out with the splitting method \citep[see][for details]{Edwards_2023b}, which accounts for persistence effects as a function of stellar brightness, scanning rate, and readout configuration. The extracted HST/WFC3 light curves present significant time-dependent systematics, including a linear long-term (``visit ramps'') and exponential short-term (``orbit ramps'') trend. We modeled these slopes together with a scaling factor between upstream and downstream spatial scans using Eq.~1 of \citet{Edwards_2023a}.

\section{Analysis}
\subsection{Photometry}
\label{sec:photo_fit}
We used the \pyorbit{} code \citep{malavolta_2016, malavolta_2018} to jointly model all available photometric observations of the three planets, with the primary goal of determining the times of inferior conjunction ($ T_\mathrm{C}$). The fit additionally included the orbital period ($P$), impact parameter ($ b$), planet-to-star radius ratio ($R_{\mathrm{p}}/R_{\star}$), eccentricity ($e$), stellar density ($\rho_{\star}$), and the limb-darkening coefficients ($q_1$, $ q_2$) as free parameters. Gaussian priors were imposed on these parameters based on the results of \citet{Lund_2019} and \cite{Grouffal_2026}. Priors on the limb-darkening coefficients were computed using the \textsc{PyLDTk} package \citep{Husser_2013, Parviainen_agrain_2015}, adopting the $q_1$, $q_2$ parameterization introduced by \citet{Kipping_2013}. For each instrument and observing campaign, an additional jitter term was included, to account for white noise, and added in quadrature to the photometric uncertainties.
We applied an over-sample factor to the K2 observations to match the overall cadence of the datasets ($\lessapprox 2$ mins). The light-curves were modeled using the the \texttt{batman} package \citep{kreidberg_2015}.

A visual inspection of the extracted K2 light curves revealed the presence of residual time-correlated systematics. To account for these effects, we modeled the residual noise using Gaussian process (GP) regression \citep{rasmussen_2006, haywood_2014}. We adopted a Matérn 3/2 kernel, implemented within the \texttt{celerite2} framework \citep{celerite2}, and fitted the GP simultaneously with the transit model to mitigate the impact of correlated noise on the inferred transit parameters.
The covariance between two data points separated by a time lag $\tau = |t_i - t_j|$ is described by
\begin{equation}
k(\tau) = \sigma_{\mathrm{GP}}^2 \left(1 + \frac{\sqrt{3}\,\tau}{\rho}\right) 
\exp\left(-\frac{\sqrt{3}\,\tau}{\rho}\right),
\end{equation}
where $\sigma_{\mathrm{GP}}$ and $\rho$ represent the amplitude and characteristic timescale of the correlated noise, respectively. We adopted broad, log-uniform priors on these hyperparameters, with $\log{\rho} \in [-3,\,3]$ and $\log_{10}(\sigma_{\mathrm{GP}}) \in [-6,\,6]$, corresponding to weakly informative priors spanning several orders of magnitude.

Following the approach of \cite{Leonardi_2024}, the global optimization of the model parameters was performed using the \pyde\footnote{\url{https://github.com/hpparvi/PyDE}}
differential evolution algorithm \citep{Storn_97, Parviainen_2016}, adopting a population of 632 different configurations ($\mathrm{n_{pop}}$) for 100\,000 generations.
The resulting solution was then used to initialize a Bayesian analysis carried out with the \texttt{emcee} package \citep{foremanmackey_2013}. We used 632 walkers for 600\,000 steps, with 350\,000 burn-in steps.

\begin{figure*}[!h]
    \sidecaption
    \centering
    \includegraphics[width=0.7\textwidth]{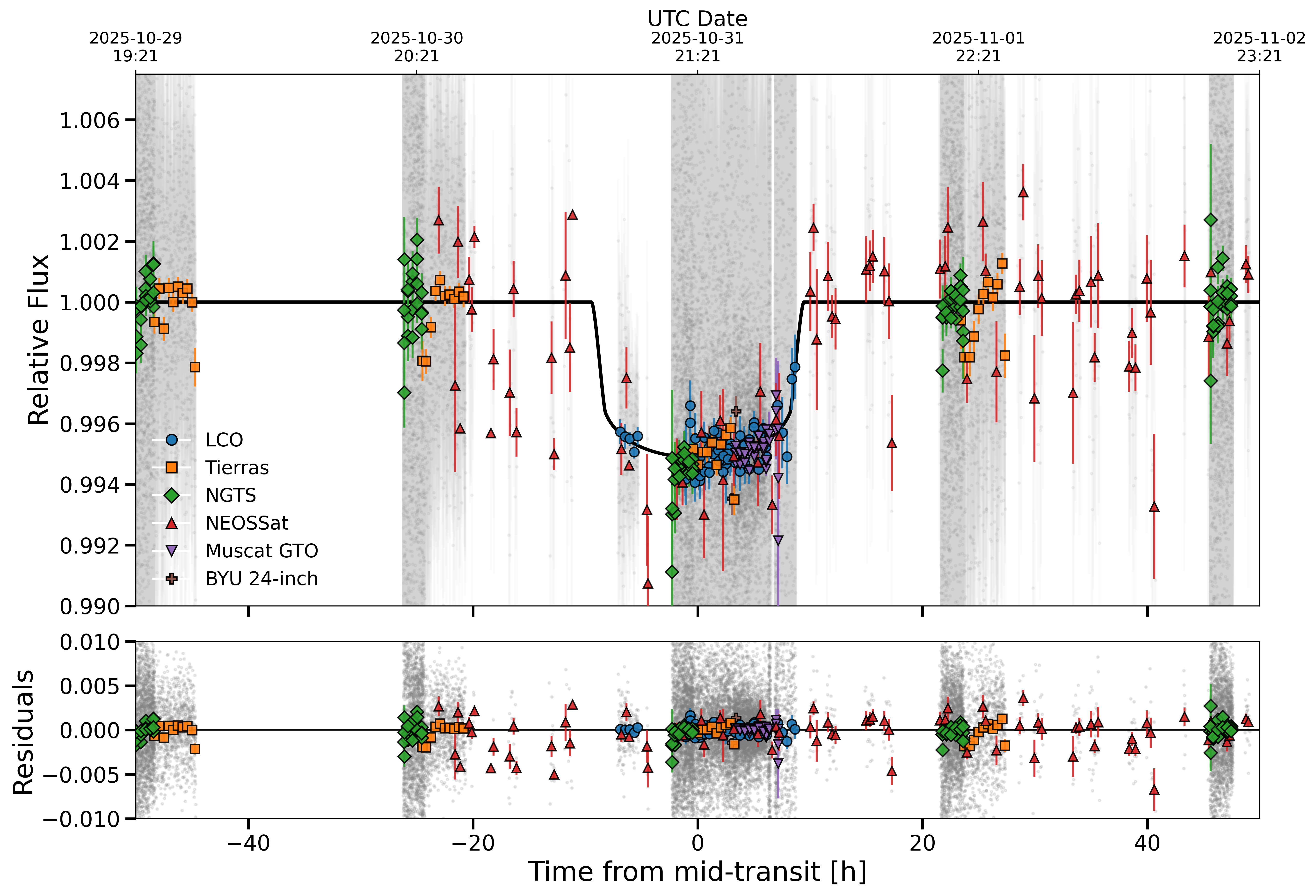}
    \caption{Combined light-curves of the 2025 transit of HIP\,41378\,f. The data from all facilities are shown together, with unbinned measurements displayed as grey points and 25-minute binned data highlighted with colored markers for each instrument. The best-fit transit model, obtained from the joint analysis with \pyorbit, is overplotted as a black line. Residuals are shown in the lower panel. The top axis indicates the corresponding UTC Date and Time.}
    \label{fig:combined_lc}
\end{figure*}

\begin{figure}
    \centering
    \includegraphics[width=1\columnwidth]{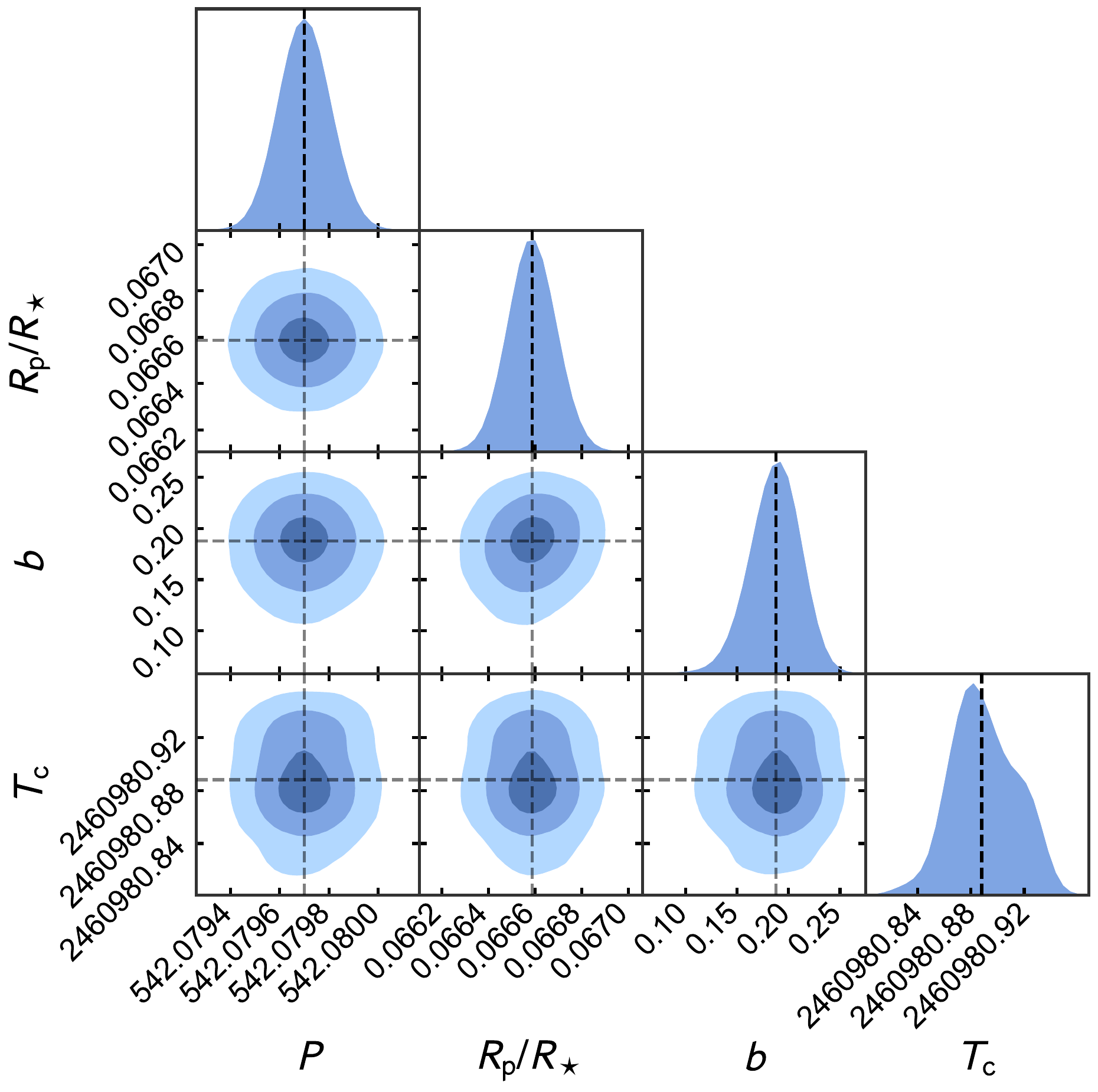}
    \caption{Posterior probability distributions for the transit parameters of HIP\,41378\,f obtained from the \pyorbit{}~\texttt{emcee} fit.}
    \label{fig:corner}
\end{figure}

\subsection{TTVs dynamical modeling}
\label{sec:dyna_fit}
During the early stages of planetary system evolution, planets can be captured into mean motion resonance (MMR) configurations, as a consequence of Type-I (non-gap-opening) convergent migration within the gas-rich proto-planetary disk \citep{malhotra_1993, kley_nelson_2012, delisle_2017, tamayo_2017, izidoro_2017, Macdonald_Dawson_2018}. In this scenario the periods of adjacent planets are near commensurability, where their ratio is close to a ratio of small integers. Resonances (or near-resonant configurations) are typically denoted as $P_1:P_2 = j:k$ MMRs, where indices 1 and 2 refer to the inner and outer planets, respectively. Such dynamical configurations can significantly amplify gravitational perturbations, leading to large variations in the times of inferior conjunction \citep{agol_2005, holman_2005}. In this framework, the TTV signal arises from the exchange of angular momentum between neighboring planets, leading to timing variations whose amplitude depends on the planetary masses and orbital eccentricities, and can range from minutes to several days \citep[e.g.][]{Barone_2025, Vitkova_2025, barkaoui_2025}.

According to the findings of \cite{Grouffal_2026}, the three outer planets in the HIP\,41378 system appear to lie close to a 3-body mean-motion resonance chain, which may explain both the observed transit timing variations of HIP\,41378\,f \citep{bryant_2021, Garcia_Meija_2025}, as well as the missed transit observation of HIP\,41378\,d \citep{sulis_2024}. For near-resonant configurations, the amplitude of TTVs increases with the orbital period of the planets, scaling linearly with $P$ \citep[eq. 5][]{agol_fabrycky_2018}. Consequently, long-period planets such as HIP\,41378\,d, e, and f are expected to exhibit particularly large timing offsets.

To account for mutual dynamical interactions among the three planets, which cannot be captured by independent Keplerian orbits, and to accurately predict future transit times, we modeled and fitted the observed transit timings using the N-body dynamical integrator \trades\footnote{\url{https://github.com/lucaborsato/trades}} \citep{borsato_2014, borsato_2019, nascimbeni_2024, borsato_2024}. The integration was initialized and run over a time span corresponding to the observational baseline ($T_\textrm{int}$ = 3840 d). All parameters were defined a common dynamical reference epoch, $T_\mathrm{ref, dyn}$ =  $2\,457\,142.0$ ($\mathrm{BJD_{TDB}}$). For each planet we put Gaussian priors on the masses, and uniform priors on the orbital periods, of planets~d and f, centered on the values retrieved from \cite{Grouffal_2026}. Given the mono-transit nature of planet e and its low significance detection in the RV data, its orbital period has yet to be fully constrained. We therefore carried out two distinct analysis. In the first configuration (1), planet e is located between planets d ($P_\mathrm{d}=278$ d) and f ($P_\mathrm{f}=542$ d), 
representing a targeted exploration of the compact architecture proposed by \cite{Grouffal_2026}, adopting a narrow uniform prior of $P_\mathrm{e} = \unif{360}{415}$. In the second scenario (2), we tested an agnostic global search across a wide parameter space, adopting a  broader uniform prior ($P_\mathrm{e} = \unif{100}{2000}$), and allowing planet e to reside either interior to planet d, between planets d and f, or exterior to planet f, thus exploring a wider range of possible orbital architectures.
The stellar radius and mass ($R_\star$, $M_\star$), as well as the planetary radii ($ R_\mathrm{p}$) and orbital inclinations $i$, were held fixed to the values reported by \cite{Lund_2019} and from our photometry analysis (see Table~\ref{tab:Transit_fit_results}). We adopted the parametrization $\left(\sqrt{e}\cos\omega, \sqrt{e}\sin\omega\right)$ in place of directly fitting $e$ and $\omega$ \citep{anderson_2011}. The longitudes of the ascending node were fixed at $\Omega = 180^\circ$ for all the planets. For each planet, the fitted parameters included the orbital period $P$, the planet-to-star mass ratio $M_\mathrm{p}/M_\star$, and the mean longitude $\lambda$. The mean longitude is defined as $\lambda = \mathcal{M} + \omega + \Omega$, where $\mathcal{M}$ denotes the mean anomaly, $\omega$ the argument of periastron, and $\Omega$ the longitude of the ascending node.

To place additional constraints on the allowed orbital periods and corresponding transit windows of planets d and e, we in corporated the extensive photometric monitoring accumulated over the past decade. In particular, for planet e we exploited the absence of detectable transit signals in Campaign 18 of the K2 mission, in CHEOPS observations from the GO program PR240015 (PI: Grouffal) and in eight TESS sectors (7, 34, 44, 45, 46, 61, 72 and 88). For planet d, we used the CHEOPS continuous observing window reported in \citep{sulis_2024}. We imposed these observing windows as exclusion constraints in the dynamical analysis, rejecting orbital solutions that would predict transit events within the covered time intervals.

After an initial exploration of the parameter space using the \pyde, with {$n_\mathrm{{pop}} = 148$} over 100\,000 generations, we sampled the posterior distributions using \emcee{} \citep{foremanmackey_2013}. The MCMC chains ({$n_\mathrm{{chains}} = 148$}) were evolved for 400\,000 steps, with the first 200\,000 steps discarded as burn-in, and a thinning factor of 100. We discarded the six burn-in values after checking convergence by Gelman-Rubin \citep{gelmanrubin_1992}, Geweke \citep{geweke_1991}, and ACF \citep{goodman_weare_2010} statistics, requiring the total chain length to exceed 50 integrated autocorrelation times. 

For each configuration we further asses the stability of the solutions by verifying the AMD–Hill stability criterion (eq.~26 of \cite{petit_2018}), which is based on the angular momentum deficit ({\tt AMD}; \citealt{laskar_1997, laskar_2000, laskar_petit_2017}), finding the entire posteriors are stable. 
From the marginalized posterior distribution, we computed high density intervals (HDI) at 68\% of each fitted parameter.
As a best-fit solution, we took the maximum a posteriori (MAP) configuration whose fitting parameters were within HDI at 68\% ($\mathrm{MAP_{HDI}}$).

\begin{figure*}[ht!]
    \centering

    \begin{subfigure}{0.49\textwidth}
        \centering
        \includegraphics[width=\linewidth]{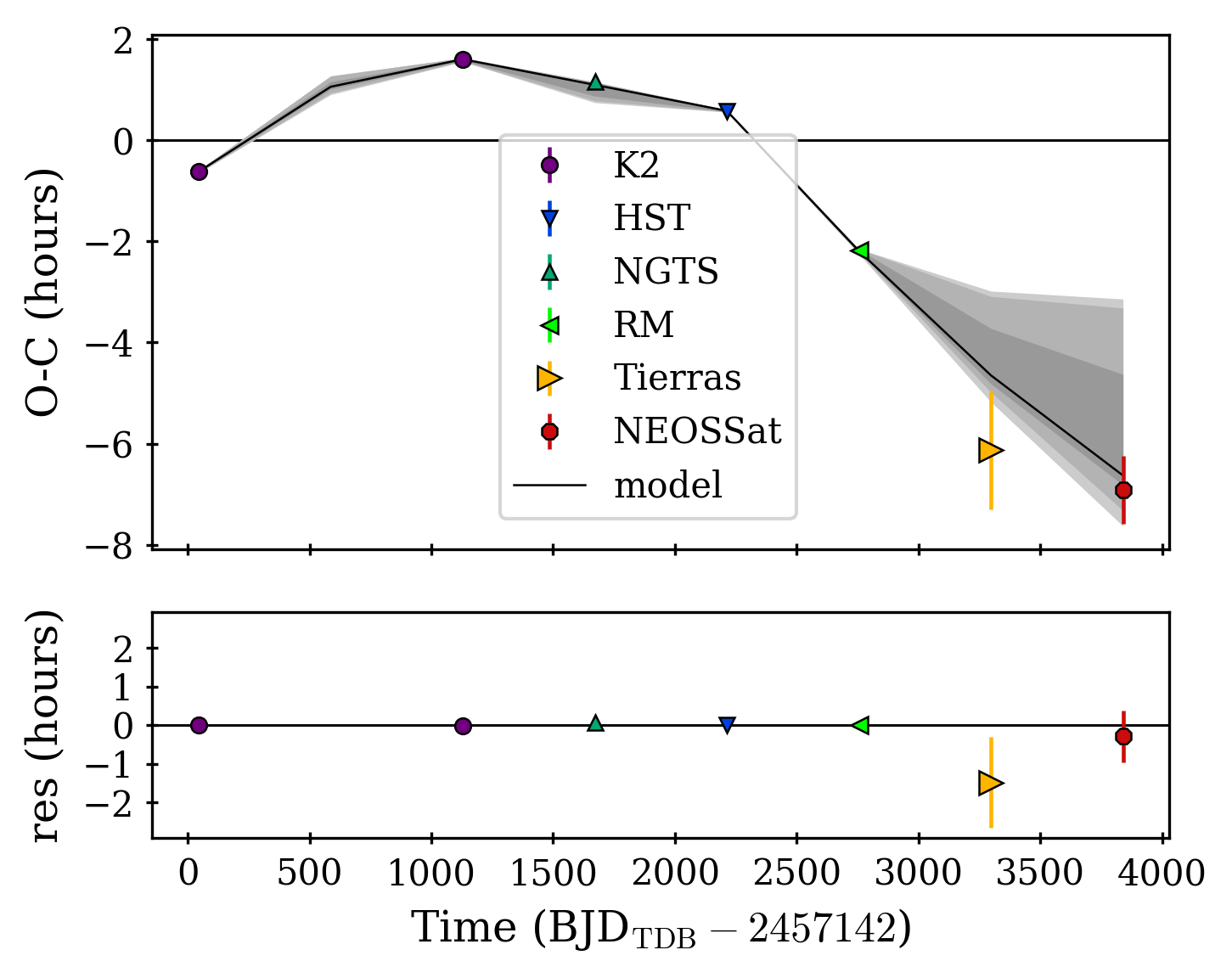}
        \caption{Configuration 1 (Compact)}
        \label{fig:oc_f_i}
    \end{subfigure}
    \hfill
    \begin{subfigure}{0.49\textwidth}
        \centering
        \includegraphics[width=\linewidth]{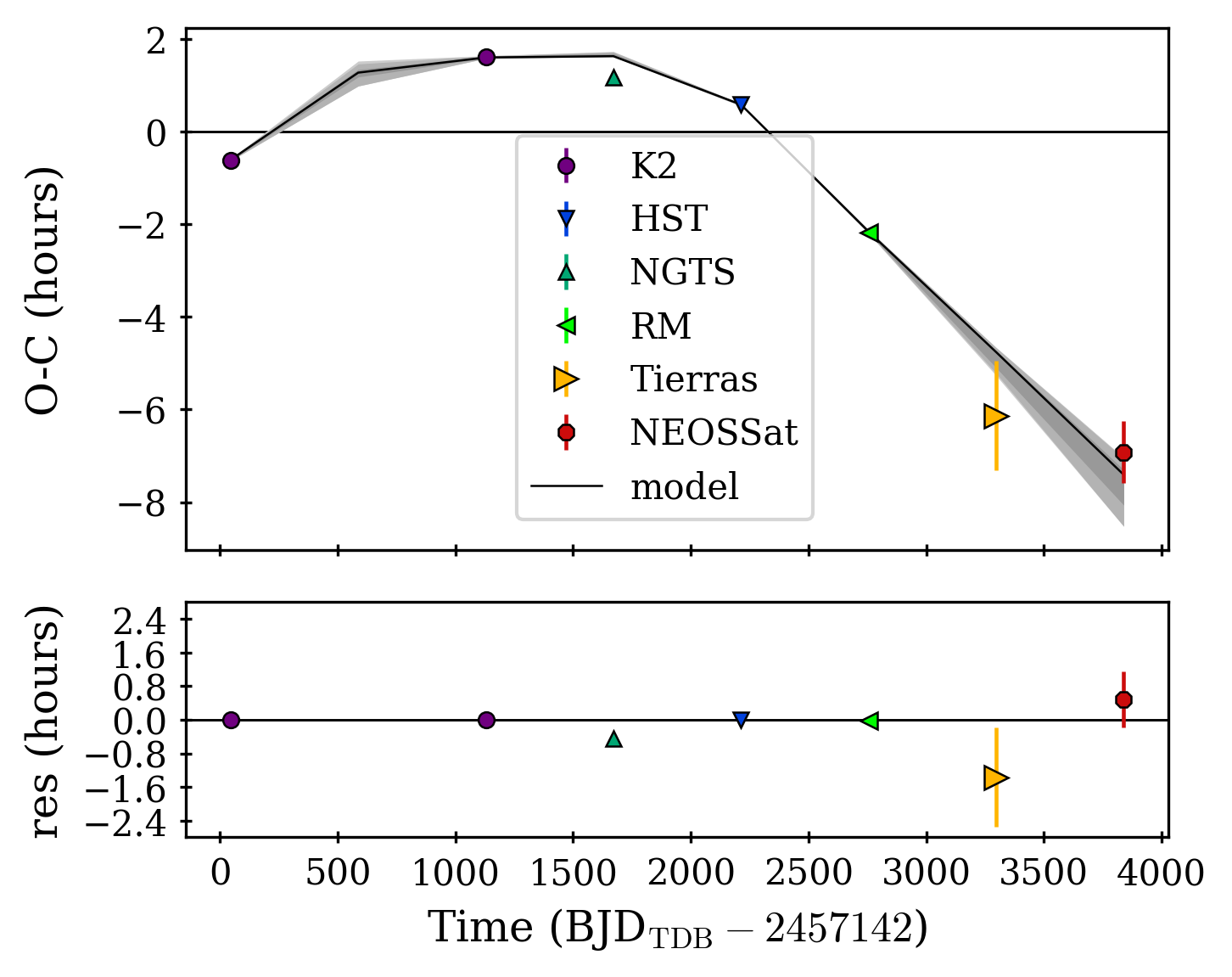}
        \caption{Configuration 2 (Extended)}
        \label{fig:oc_f_ii}
    \end{subfigure}

    \caption{Observed-minus-calculated ($O-C$) diagrams for HIP\,41378\,f for the compact configuration (a) and the extended configuration (b). The best-fit model ($\mathrm{MAP_{HDI}}$) is shown as a black curve, while the shaded regions represent the 1, 2, and 3$\sigma$ posterior intervals}.    
    \label{fig:oc_f}
\end{figure*}

\section{Results and discussions}
\label{sec:resu}

\subsection{Photometric results}

By jointly modeling the newly acquired light curves of HIP\,41378\,f together with archival observations, we derived a precise measurement of its 2025 transit time, $T_{\mathrm{C},7} = 2\,460\,980.888~\pm ~0.029~\mathrm{BJD_{TDB}}$. This value is consistent within $1\sigma$ with the prediction of \cite{Garcia_Meija_2025}. The 2025 combined transit light curve of HIP\,41378\,f is shown in Fig.~\ref{fig:combined_lc}, while the corresponding posterior distributions are presented in Fig.~\ref{fig:corner}. In addition, we re-extract the transit timings for all three planets from the available K2 and HST datasets using a homogeneous analysis (see Fig.~\ref{fig:K2_HST_lcs}). All retrieved planetary parameters and transit times, together with literature $T_{C}$ values (from \citealt{bryant_2021, grouffal_2022} and \citealt{Garcia_Meija_2025}), are reported in Tables~\ref{tab:Transit_fit_results} and~\ref{tab:T0s}.
Overall, the derived orbital and planetary parameters are consistent with those reported by \cite{Grouffal_2026}. However, our analysis does not provide meaningful constraints on the orbital eccentricities of the three planets. This is expected, as eccentricity is weakly constrained by transit photometry alone due to strong degeneracies with other transit parameters (e.g., impact parameter and stellar density). Consequently, we report 2$\sigma$ (95\% confidence) upper limits for these parameters (see Table~\ref{tab:Transit_fit_results}).


\begin{table*}
\centering
\renewcommand{\arraystretch}{1.3}
\caption{Posteriors and derived orbital parameters for HIP\,41378\,d, e, and f obtained from the dynamical analysis with \trades.}
\begin{tabular}{l l c c c c}
\hline\hline
Planet & Parameter & Unit & Prior & Configuration 1 (Compact) & Configuration 2 (Extended) \\
\hline

HIP\,41378\,d 
& Orbital Period & [d] 
& \unif{277}{279} 
& $278.3272_{-0.0091}^{+0.0172}$ 
& $278.587_{-0.015}^{+0.016}$ \\

& Mass & [$M_{\oplus}$] 
& $\mathcal{G}(6.53^{+2.65}_{-3.12})$ 
& $9.52_{-0.31}^{+2.87}$ 
& $3.37_{-0.34}^{+0.66}$ \\

& Eccentricity & 
& \unif{0}{0.5} 
& $0.1391_{-0.0147}^{+0.0047}$ 
& $0.1321_{-0.0100}^{+0.0060}$ \\

\hline

HIP\,41378\,e 
& Orbital Period & [d] 
& See note\tablefootmark{a} 
& $389.12_{-0.28}^{+0.16}$ 
& $993_{-3}^{+2}$ \\

& Mass & [$M_{\oplus}$] 
& $\mathcal{G}(7.62^{+4.63}_{-3.12})$ 
& $6.18_{-0.46}^{+1.10}$ 
& $14_{-2}^{+1}$ \\

& Eccentricity & 
& \unif{0}{0.5} 
& $0.0236_{-0.0028}^{+0.0046}$ 
& $0.20856_{-0.01482}^{+0.00095}$ \\

\hline

HIP\,41378\,f 
& Orbital Period & [d] 
& \unif{541}{543} 
& $542.2484_{-0.0052}^{+0.0342}$ 
& $541.756_{-0.015}^{+0.051}$ \\

& Mass & [$M_{\oplus}$] 
& $\mathcal{G}(25.0^{+5.0}_{-5.0})$ 
& $25.0_{-3.0}^{+5.0}$ 
& $33.0_{-3.0}^{+3.0}$ \\

& Eccentricity & 
& \unif{0}{0.5} 
& $0.0140_{-0.0053}^{+0.0021}$ 
& $0.00098_{-0.00066}^{+0.00111}$ \\

\hline
\end{tabular}

\label{table:TRADES_results}

\tablefoot{
\tablefoottext{a}{For Configuration 1, the prior on $P_e$ is $\mathcal{U}(360, 415)$~days, while for Configuration 2 a broader prior of $\mathcal{U}(100, 2000)$~days is adopted.} The symbols $\mathcal{U}(a,b)$ and $\mathcal{G}(\mu^{+\sigma_1}_{-\sigma_2})$ denote uniform and Gaussian priors, respectively. 
For the uniform prior, $a$ and $b$ represent the lower and upper bounds. For the Gaussian prior, $\mu$ is the mean, while $\sigma_1$ and $\sigma_2$ correspond to the upper and lower uncertainties, respectively, allowing for asymmetric dispersions.}
\end{table*}

\subsection{TTVs modeling results}
The derived transit timings were used to investigate the dynamical architecture of the system through N-body modeling.
We explored two dynamical configurations consistent with the available constraints. The derived parameters of both analyses can be found in Table~\ref{table:TRADES_results} and the O-C diagrams of HIP\,41378\,f are shown in Fig.~\ref{fig:oc_f}.
Given the limited number of available transit timing measurements (10) compared to the number of fitted parameters (15), a formal statistical comparison between the two configurations is not warranted. In this regime, model selection metrics such as Bayesian evidence or information criteria (e.g. BIC or AIC) are known to be unreliable \citep{kass_95, Burnham_Anderson_2002, Johnson_Lebront_2004, Liddle_2007, Trotta_2008}. We therefore refrain from favoring one configuration over the other and present both solutions.

Likewise, while both configurations formally lead to an improved precision on the planetary masses, these constraints are likely driven by the adopted Gaussian priors and the limited number of transit timings. We therefore caution against over-interpreting the derived masses. Instead, our analysis is primarily aimed at exploring the dynamical interactions within the system and at predicting future transit windows (see Sect.~\ref{sec:pred}).


In the first configuration, we follow the results of \cite{Grouffal_2026}, whose more than a decade of RV monitoring suggests a compact dynamical architecture in which planet e is located between planets d and f. This scenario yields a best-fit orbital period of $P_e = 389.12_{-0.28}^{+0.16}$ d, placing the three planets near a chain of second-order 7:5 two-body mean-motion resonances. 

High-order MMRs (e.g., 5:3, 7:5) are generally expected to be less frequent and dynamically weaker than first-order ones. This is because standard convergent migration typically favors trapping into stronger, lower-order commensurabilities (e.g., 3:2, 2:1) before planets can reach closer configurations \citep{dai_2024}. However, capture in second and third-order MMRs, during Type-I migration, is not dynamically forbidden \citep{keller_2026}, for instance, theory predicts that these tight period ratio spacings are more common in lower-density disks \citep{xu_lai_2017}.
In addition, the strength of the resonant terms scales as $\propto e^{|j-k|}$, where $|j-k|$ defines the order of the resonance, implying that non-zero eccentricities are required to activate the corresponding terms in the disturbing function and lead to significant TTVs modulation.

The best-fit solution yields a moderately eccentric orbit for planet d ($e_\mathrm{d} = 0.1391_{-0.0147}^{+0.0047}$). This value is consistent with our 95\% confidence upper limit (see Table.~\ref{tab:Transit_fit_results}), but differs from the results of \cite{Grouffal_2026}, who report low ($<0.1$) eccentricities for all three planets. This discrepancy may arise from the use of a truncated normal prior ($\mathcal{N}(0.0,\,0.083)$) in their analysis, which could have biased the posterior distribution toward lower eccentricity values.

In the second configuration, we relaxed the boundary on the orbital period of HIP\,41378\,e, allowing it to vary over a wide range ($P_\mathrm{e} \in [100, 2000]$ d). This yields a best-fit solution of $P_\mathrm{e} = 993_{-3}^{+2}$ d, corresponding to a different dynamical architecture in which the system approaches a 4:2:1 three-body commensurability. In this scenario, HIP\,41378\,d and f lie close to a 2:1 MMR ($\Delta \sim -2.7\%$), while HIP\,41378\,e is near a 2:1 commensurability with planet f ($\Delta \sim -8.3\%$).

The sampler in Configuration 2 converged exclusively toward the extended solution around $P_e \sim 991$~days, completely bypassing the compact solution. This confirms that the two configurations represent genuinely independent, isolated dynamical islands in the multi-dimensional parameter space.

In both configurations, the posterior distributions favor moderately eccentric orbits, particularly for planets d and e (see Table~\ref{table:TRADES_results}), consistent with our 95\% confidence upper limits. This is broadly in line with population-level studies showing that long-period planets tend to retain higher eccentricities, as tidal circularization becomes inefficient at large orbital separations \citep[e.g.,][]{jackson_2008, wright_2009}. In this regime, orbital eccentricities are primarily shaped by dynamical processes such as planet--planet scattering and secular interactions \citep{rasio_1996, chatterjee_2008, juric_tremaine_2008, naoz_2016}. 

For both configurations, the best-fit TTV models for planet f (Fig.~\ref{fig:oc_f}) are nearly indistinguishable. However, the extended (2) configuration fails to accurately reproduce the transit time reported by \cite{bryant_2021}, predicting a transit occurring $\sim$0.5--1 hour later than observed. Given the precision of that measurement, this discrepancy suggests tension between the extended configuration and the available data.

Additional photometric and spectroscopic observations, together with future joint analyses combining TTVs and radial velocities, will be essential to further constrain the system architecture. This is however beyond the scope of the present work and will be addressed in a forthcoming study.

\begin{table*}[t]
\centering
\renewcommand{\arraystretch}{1.3}
\caption{Predicted future transit times for HIP\,41378\,d, e, and f.}
\begin{tabular}{l l c c c c c}
\hline\hline
Configuration & Planet &
$T_\mathrm{ref}$ &
$P$ &
Predicted $T_0$ &
Date (UTC) \\
&
&
(BJD$_\mathrm{TDB}$) &
(d) &
(BJD$_\mathrm{TDB}$) &
 \\
\hline

\multirow{4}{*}{(1) Compact}
& HIP\,41378\,d &
$2\,459\,394.3740 \pm 0.1057$ &
$278.6144 \pm 0.0228$ &
$2\,461\,066.53 \pm 0.35$ &
26-01-2026 \\
&
&
&
&
$2\,461\,345.35 \pm 0.39$ &
31-10-2026 \\

& HIP\,41378\,e &
$2\,459\,475.7472 \pm 0.1286$ &
$388.8286 \pm 0.0369$ &
$2\,461\,419.25 \pm 2.97$ &
13-01-2027 \\

& HIP\,41378\,f &
$2\,459\,355.0009 \pm 0.0303$ &
$541.9859 \pm 0.0117$ &
$2\,461\,522.86 \pm 0.07$ &
27-04-2027 \\

\hline

\multirow{4}{*}{(2) Extended}
& HIP\,41378\,d &
$2\,459\,394.0952 \pm 0.1146$ &
$278.5603 \pm 0.0247$ &
$2\,461\,065.95 \pm 0.14$ &
25-01-2026 \\
&
&
&
&
$2\,461\,344.80 \pm 0.17$ &
31-10-2026 \\

& HIP\,41378\,e &
$2\,459\,354.9776 \pm 0.0350$ &
$541.9751 \pm 0.0135$ &
$2\,461\,107.3 \pm 10.0$ &
07-03-2026 \\

& HIP\,41378\,f &
$2\,459\,124.5571 \pm 0.0824$ &
$991.3611 \pm 0.0583$ &
$2\,461\,522.75 \pm 0.06$ &
27-04-2027 \\

\hline
\end{tabular}
\label{tab:future_T0s}
\tablefoot{Mid-transit times ($T_0$) and their associated 1$\sigma$ uncertainties ($\sigma_{T_{0}}$) are derived from the forward propagation of the $\mathrm{MAP_{HDI}}$ for two distinct dynamical scenarios. The reference linear ephemerides ($T_\mathrm{ref}$ and $P$) applied for each planet are reported.}
\end{table*}   
\subsection{Future transits windows}
\label{sec:pred}

As shown in the Sec.\ref{sec:dyna_fit}, the limited number of transit timings available for the three planets prevents a statistically robust choice among the different dynamical models explored for the TTV analysis.
To facilitate future follow-up observations aimed at improving the dynamical characterization of the system, we performed a forward-modeling analysis of the observed O–C residuals. We propagated the $\mathrm{MAP_{HDI}}$ solution for 1.5 years beyond the last observed epoch, to cover the next potential transit opportunities of the outer planets. 
Along with the $\mathrm{MAP_{HDI}}$ solution we computed the associated HDI at 68\% and 95\% of 10\,000 random samples drawn from the posterior distribution of the two analysis.
As shown in Figure~\ref{fig:oc_predicted}, the predicted TTV signal remains coherent over the explored time span, while the uncertainty on the transit times ($\sigma_{T_\mathrm{c}}$) gradually increases. Although the current dataset does not allow us to discriminate between the explored dynamical configurations (see Sect.~\ref{sec:dyna_fit}), the forward-modeling results demonstrate that the ephemerides are sufficiently constrained to enable efficient transit recovery in the near future (see Table~\ref{tab:future_T0s}).

In particular, we precisely predict the next transit windows of HIP\,41378\,f and d as high-priority targets for follow-up observations. Given the long transit duration of both planets ($T_{14}\sim$19 and $\sim$ 12 hours) and the growing timing uncertainties, a continuous monitoring strategy is essential to ensure full or near-complete coverage of the transit event. This strongly favors a coordinated strategy combining ground-based networks (e.g., longitudinally distributed facilities) with space-based facilities, which are not affected by weather interruptions, variable seeing conditions, or the diurnal cycle, such as CHEOPS, NEOSSat, or JWST.

\section{Conclusions}
\label{sec:conclusions}

In this work we present the results of our photometric and dynamical analysis of the outer planets in the HIP\,41378 system, combining a new transit observation of HIP\,41378\,f, between 31 October 2025 and 1 November ($T_{\mathrm{C},7} = 2\,460\,980.888~\pm ~0.029~\mathrm{BJD_{TDB}}$), with re-extracted transit timings of HIP\,41378\,d, e and f. Using this updated set of transit times, we performed N-body simulations to model the dynamical evolution of the outer subsystem. Leveraging a timing baseline spanning more than a decade, we refine the ephemerides of the three planets and provide updated predictions for future transit events, enabling efficient follow-up observations.

\begin{figure*}[ht!]
    \centering

    \begin{subfigure}{0.49\textwidth}
        \centering
        \includegraphics[width=\linewidth]{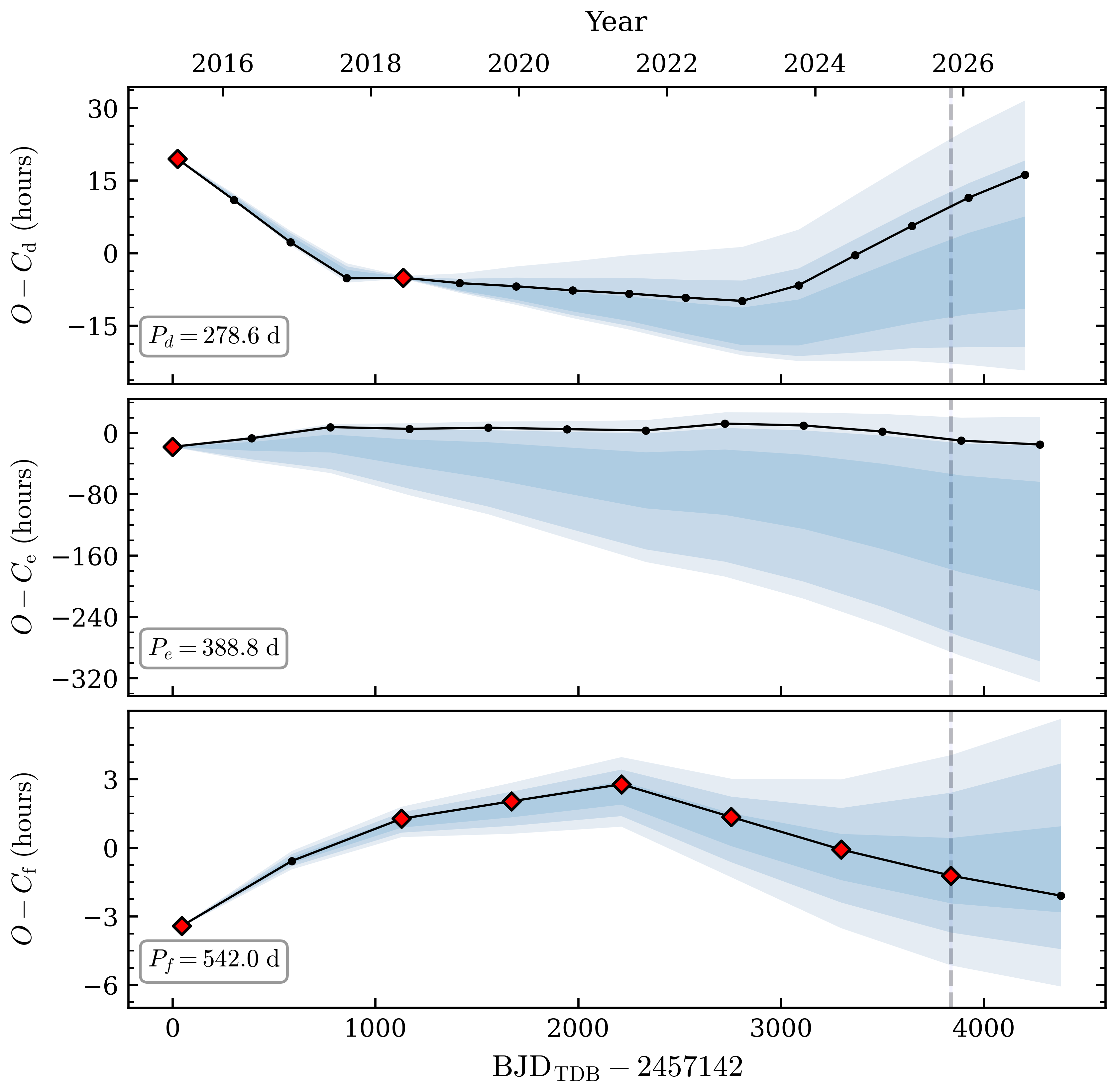}
        \caption{Configuration 1 (Compact)}
        \label{fig:oc_i}
    \end{subfigure}
    \hfill
    \begin{subfigure}{0.49\textwidth}
        \centering
        \includegraphics[width=\linewidth]{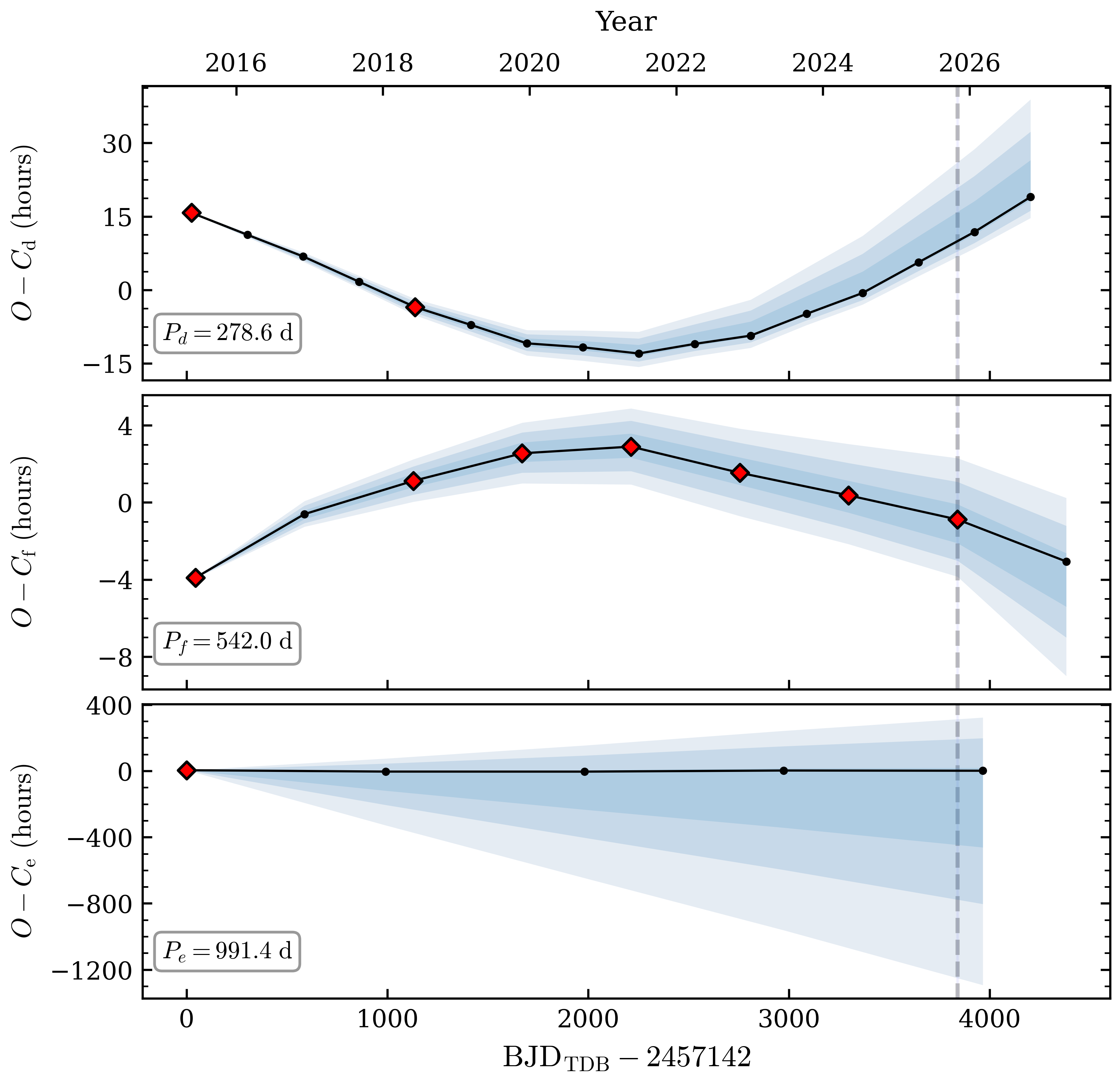}
        \caption{Configuration 2 (Extended)}
        \label{fig:oc_ii}
    \end{subfigure}

    \caption{Forward modeling of the $O-C$ diagrams for the HIP\,41378 outer triplet based on the $\mathrm{MAP_{HDI}}$ solution. The top axis shows the calendar year of observation, while the bottom axis gives the time in $\mathrm{BJD_{TDB}}$. Black points represent the simulated individual transit times, and the solid black line shows the best-fit dynamical model. Red diamonds mark the observed epochs for each planet, while the blue shaded regions indicate the $1\sigma$, $2\sigma$, and $3\sigma$ confidence intervals. Panel (a) shows the TTV signal for configuration~1, and panel (b) for configuration~2. The vertical dashed lines indicate the end of the current observational baseline. The O-C values are computed with respect to the linear ephemeris reported in Table~\ref{tab:future_T0s}.}    \label{fig:oc_predicted}
\end{figure*}

Our analysis did not allow to resolve the dynamical architecture of the outer system, with two configurations providing comparably good fits to the current TTV dataset. Despite this, the TTV signals clearly indicate strong dynamical coupling between the three giant planets, as illustrated in the Figs.~\ref{fig:oc_i},~\ref{fig:oc_ii}. This system therefore represents a valuable laboratory for studying long-period, dynamically interacting exoplanets, a regime largely unexplored so far due to the observational limitations.

Considering the loose constraints on the predicted transit windows of HIP\,41378\,e and the absence of a significant Keplerian reflex signal in more than a decade of radial-velocity observations \citep{Grouffal_2026}, detecting additional transits of this planet will be challenging. In this context, refining our knowledge of this planet may instead rely on follow-up observations of its neighboring planetary companions.
Additional transit detections of HIP\,41378\,d and f could significantly improve the retrieval of planet e period, rapidly reducing the ephemeris uncertainties.
Even a single new timing measurement would provide strong leverage to distinguish between the proposed dynamical configurations, given their diverging TTV predictions on relatively short timescales.

\begin{acknowledgements}
We thank the anonymous for the insightful and constructive comments, which significantly improved the clarity and focus of the manuscript.
PLe, LBo and GPi acknowledge support from CHEOPS ASI-INAF agreement n. 2019-29-HH.0.
LBo acknowledges financial support from the Bando Ricerca Fondamentale INAF 2023, Mini-Grant: "Decoding the dynamical properties of planetary systems observed by TESS and CHEOPS through TTV analysis with parallel computing". The project leading to this publication has received funding from the Excellence Initiative of Aix-Marseille University - A*Midex, a French “Investissements d’Avenir programme” AMX-19-IET-013. This work was supported by the "Programme National de Planétologie" (PNP) of CNRS/INSU co-funded by CNES. This work was supported by the "Action Thématique Physique Stellaire” of CNRS/INSU PN ASTRO co-funded by CEA and CNES.
This work makes use of observations from the Las Cumbres Observatory global telescope network. Part of the LCOGT telescope time was granted by NOIRLab through the Mid-Scale Innovations Program (MSIP). MSIP is funded by NSF. Part of the LCOGT time is based on observations made with the Las Cumbres Observatory's education network telescopes that were upgraded through generous support from the Gordon and Betty Moore Foundation. One LCOGT observation was conducted with time provided to Boyce Research Initiatives and Education Foundation by the Las Cumbres Observatory through its Global Sky Partners program. All other LCOGT observations were conducted with time provided to the TESS Follow-up LCOGT Key Project (PI: A. Shporer).

This research has made use of the Exoplanet Follow-up Observation Program (ExoFOP; DOI: 10.26134/ExoFOP5) website, which is operated by the California Institute of Technology, under contract with the National Aeronautics and Space Administration under the Exoplanet Exploration Program.

Funding for the TESS mission is provided by NASA's Science Mission Directorate. KCo acknowledges support from the TESS mission via NASA grants 80NSSC24K1889 and 80NSSC26K0081.

This paper is based on observations made with the MuSCAT3 instrument, developed by the Astrobiology Center and under financial supports by JSPS KAKENHI (JP18H05439) and JST PRESTO (JPMJPR1775), at Faulkes Telescope North on Maui, HI, operated by the Las Cumbres Observatory. This work is partly supported by JSPS KAKENHI Grant Numbers JP24H00017, JP24K00689, JSPS Bilateral Program Number JPJSBP120249910, JST SPRING, Grant Number JPMJSP2108 and JSPS Grant-in-Aid for JSPS Fellows Grant Number JP25KJ1036. 
Based on data collected under the NGTS project at the ESO La Silla Paranal Observatory. The NGTS facility is operated by the consortium institutes with support from the UK Science and Technology Facilities Council (STFC)  projects ST/M001962/1 and  ST/S002642/1.
This work has made use of data from the European Space Agency (ESA) mission {\it Gaia} (\url{https://www.cosmos.esa.int/gaia}), processed by the {\it Gaia}
Data Processing and Analysis Consortium (DPAC, \url{https://www.cosmos.esa.int/web/gaia/dpac/consortium}). Funding for the DPAC has been provided by national institutions, in particular the institutions participating in the {\it Gaia} Multilateral Agreement.
JSj gratefully acknowledges support by FONDECYT grant 1240738 and from the ANID BASAL project FB210003. The Tierras Observatory is supported by the National Science Foundation under Award No. AST-2308043. JGm acknowledges support from the Heising-Simons Foundation, and from the Pappalardo family through the MIT Pappalardo Fellowship in Physics.
\end{acknowledgements}

\bibliographystyle{aa}
\bibliography{references}

\begin{appendix}
\label{appendix}
\clearpage

\clearpage
\onecolumn

\section{Additional plots}

\begin{figure*}[!h]
    \centering
    \begin{tabular}{cc}
        \includegraphics[width=0.35\textwidth]{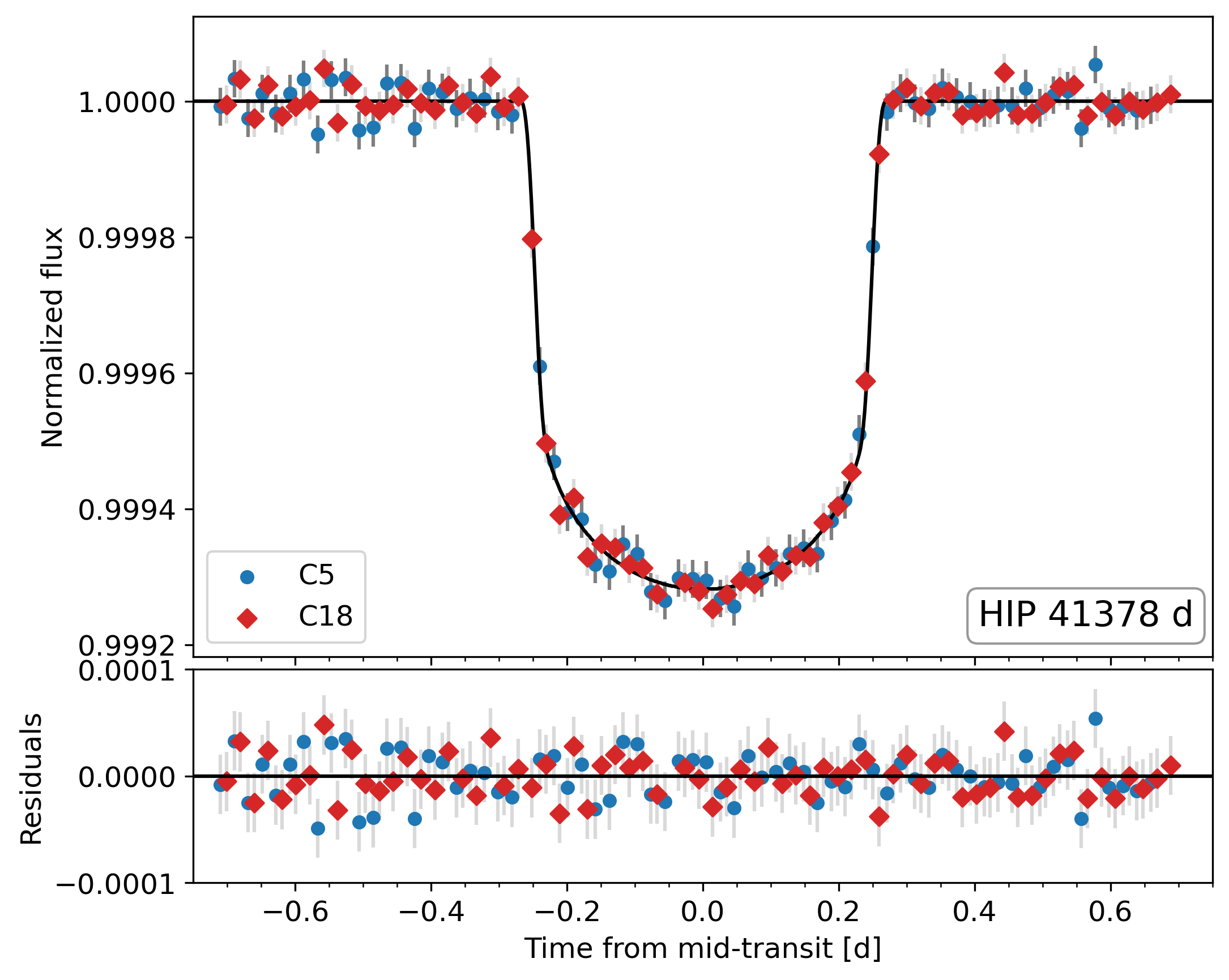} & 
        \includegraphics[width=0.35\textwidth]{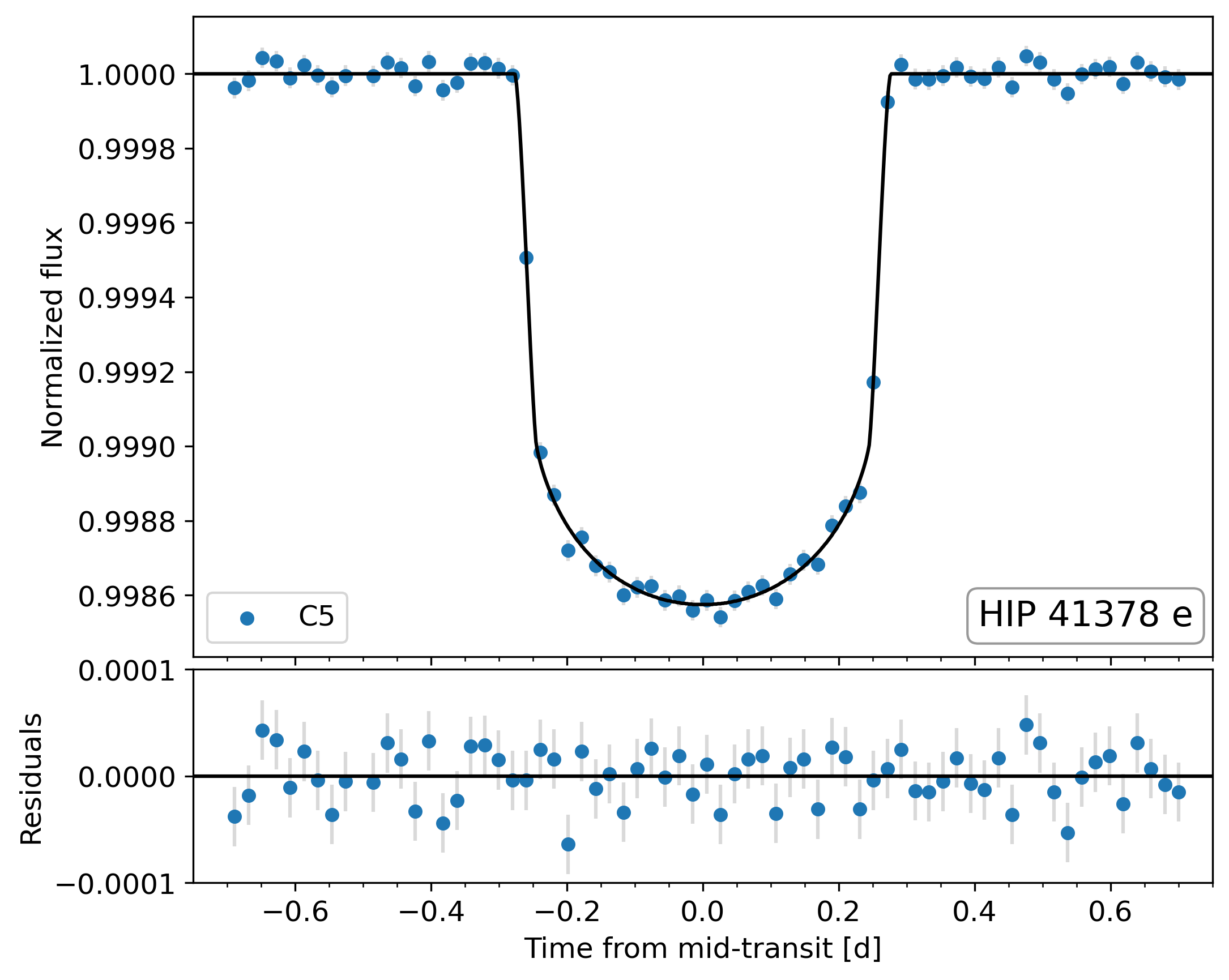} \\
        \noalign{\smallskip} 
        \includegraphics[width=0.35\textwidth]{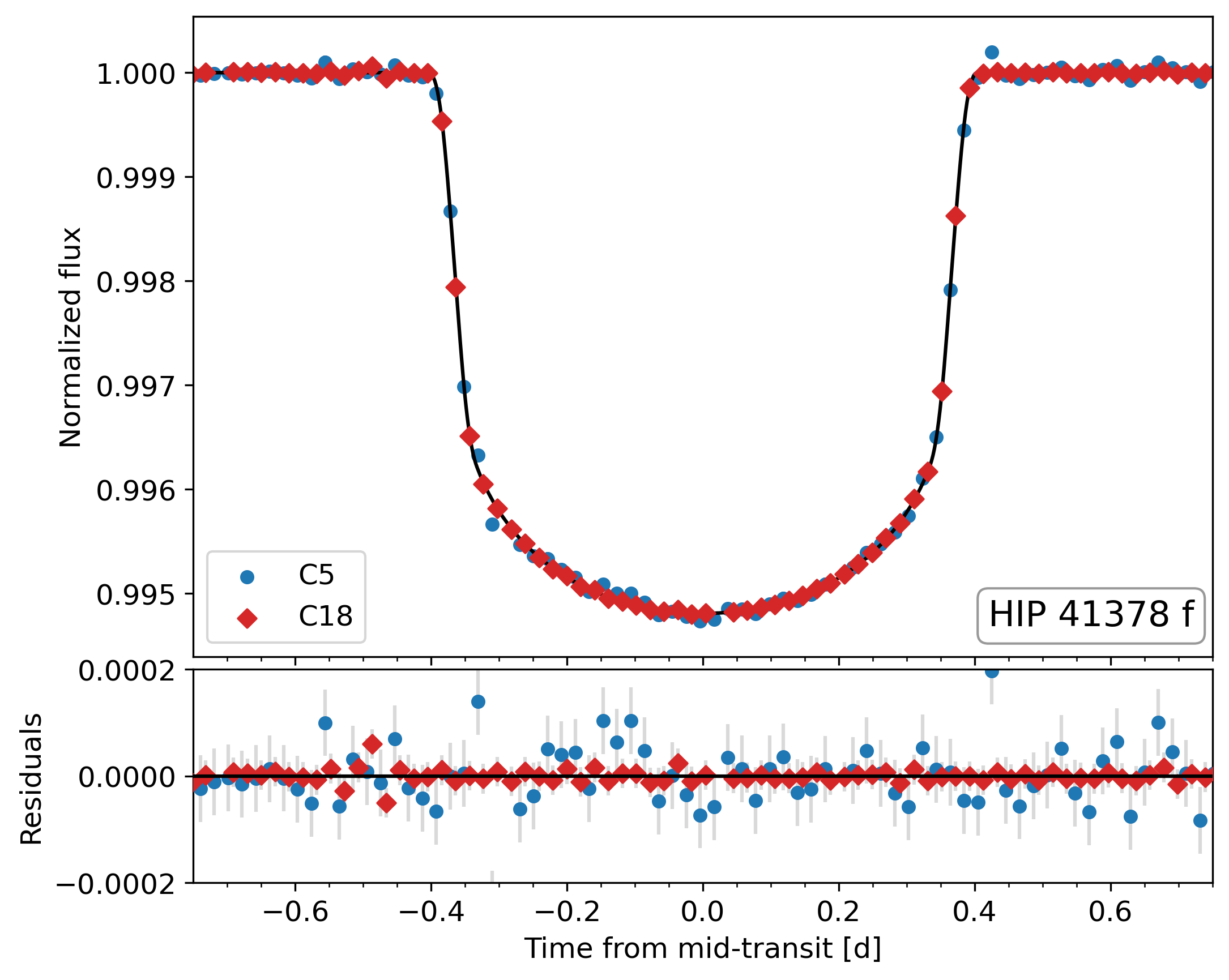} & 
        \includegraphics[width=0.35\textwidth]{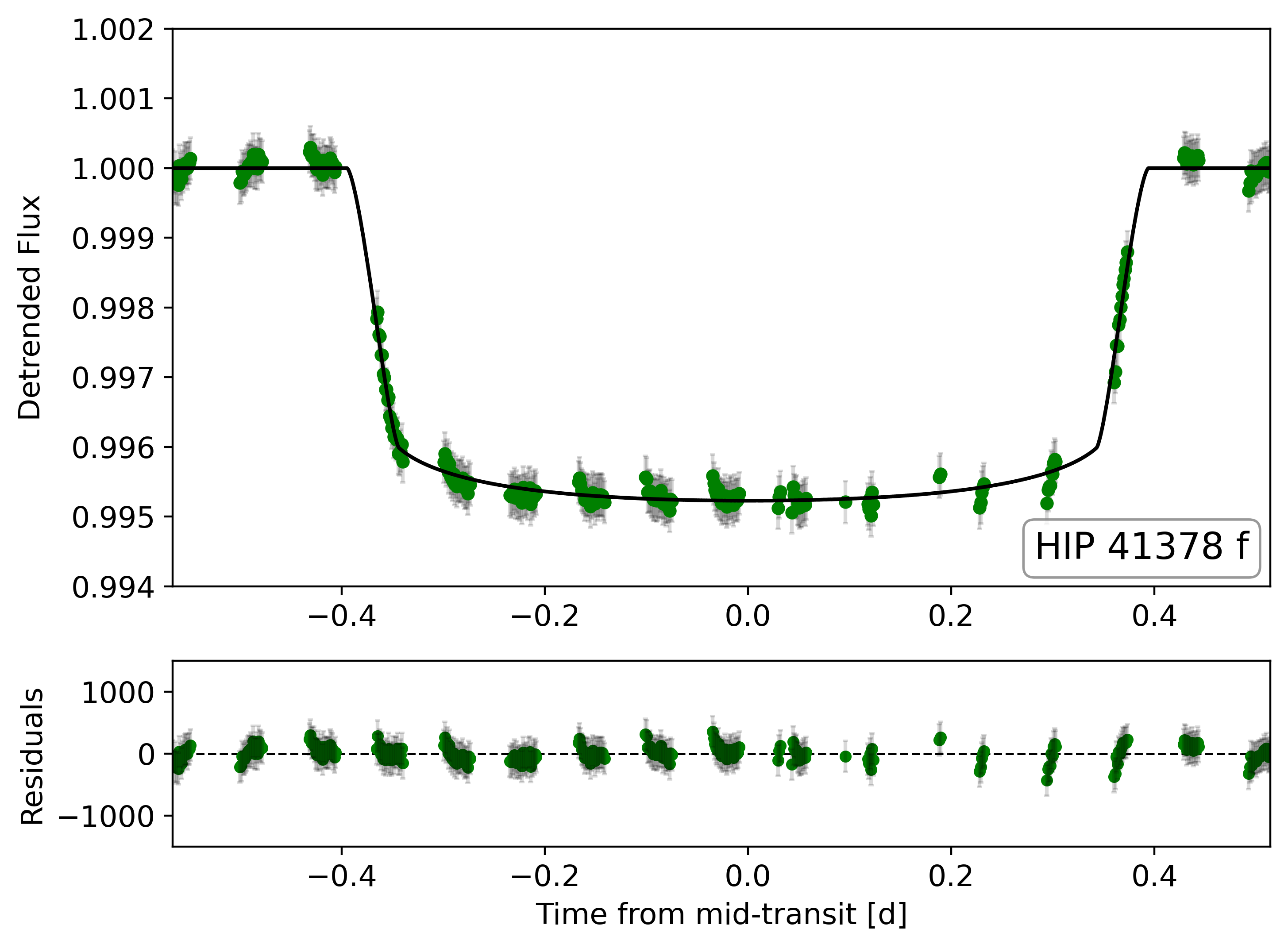} \\
    \end{tabular}
    \caption{Transit light curves of HIP\,41378\,d, e, and f from K2 Campaign 5 (C5; blue circles) and Campaign 18 (C18; red diamonds), and HST (green points). All light curves are centered on their respective mid-transit times. The black solid line shows the best-fit transit model obtained from a joint fit of all datasets performed with \texttt{PyORBIT}. Residuals are shown in the lower panels.}
    \label{fig:K2_HST_lcs}
\end{figure*}

\begin{figure}[!h]
   \centering
    \includegraphics[width=0.7\textwidth]{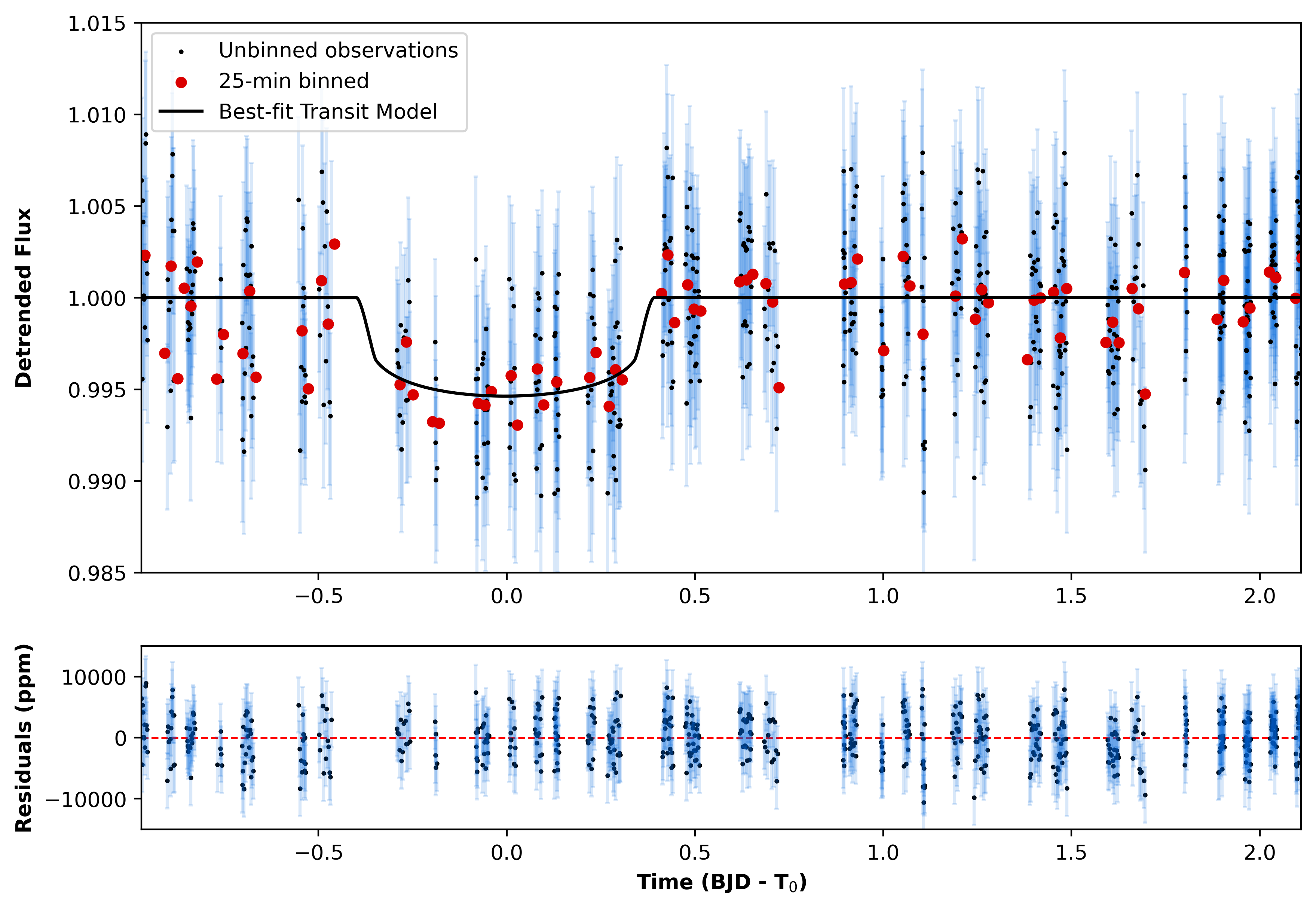}
    \caption{NEOSSat 2025 transit light curve of HIP\,41378\,f. \textit{Top:} Detrended photometric data are shown as black points with blue error bars, while the 25-min binned measurements are highlighted in red. The best-fitting transit model is overplotted as a black solid line. \textit{Bottom:} Residuals of the fit.}

    \label{fig:lc_neossat}
\end{figure}

\begin{figure}
    \centering
    \includegraphics[width=1\textwidth]{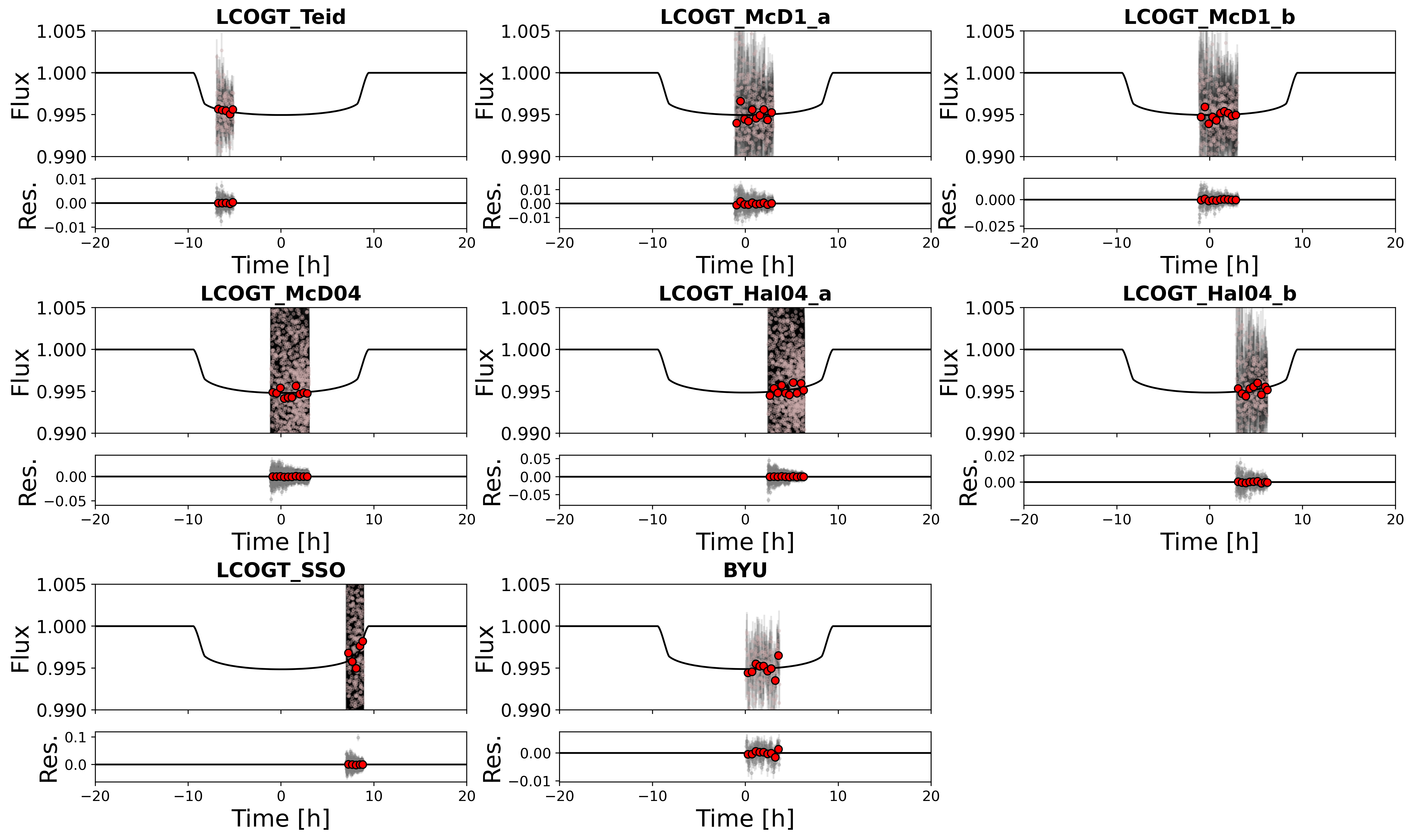}
    \caption{Same as Fig.\ref{fig:lc_neossat} for LCOGT and BYU 24 inch telescopes light-curves.}
    \label{fig:LCO_lcs}
\end{figure}


\begin{figure}
    \centering
    \includegraphics[width=1\textwidth]{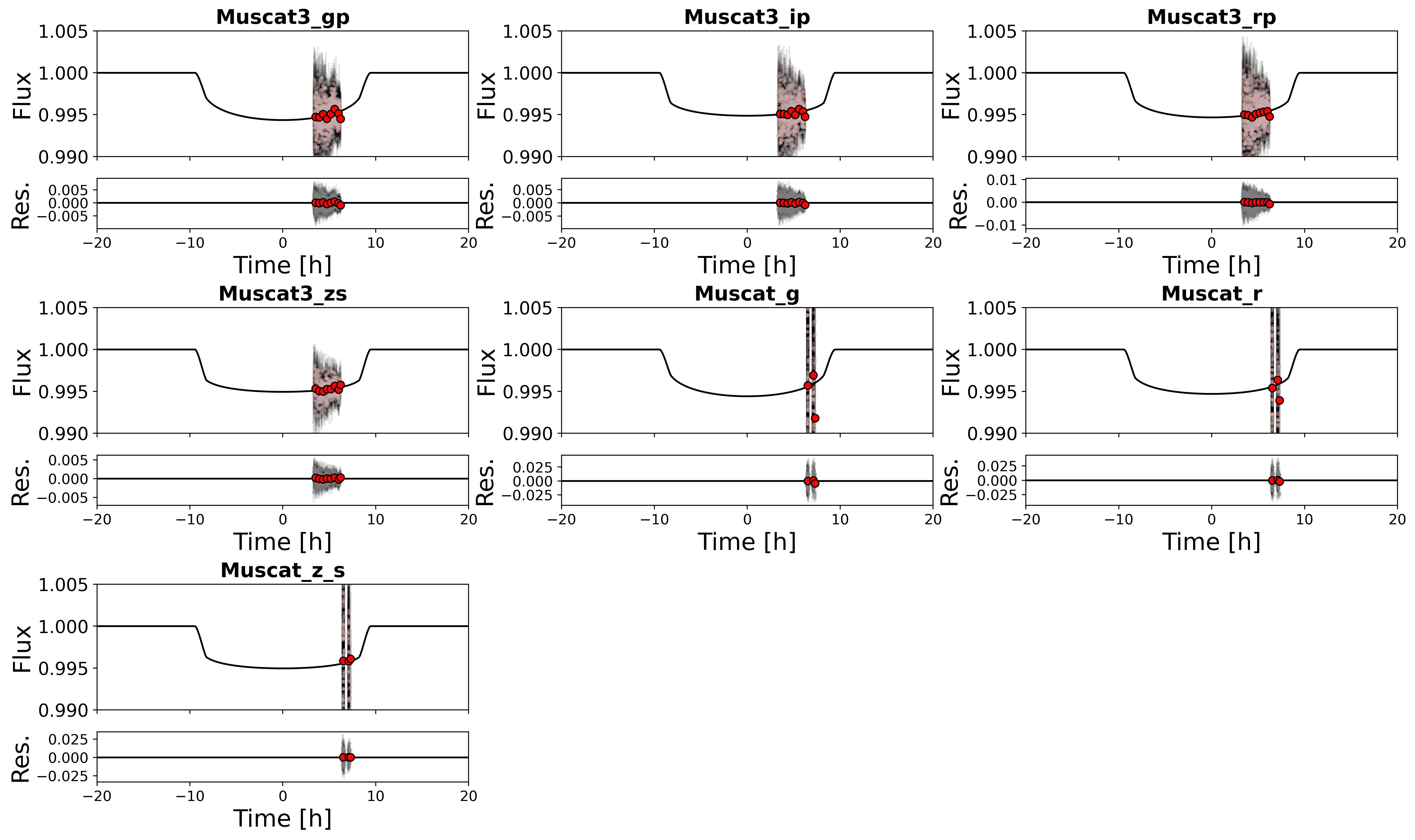}
    \caption{Same as Fig.\ref{fig:lc_neossat} for MuSCAT and MuSCAT3 ligh-curves.}
    \label{fig:Muscat_lcs}
\end{figure}


\begin{figure}
    \centering
    \includegraphics[width=1\textwidth]{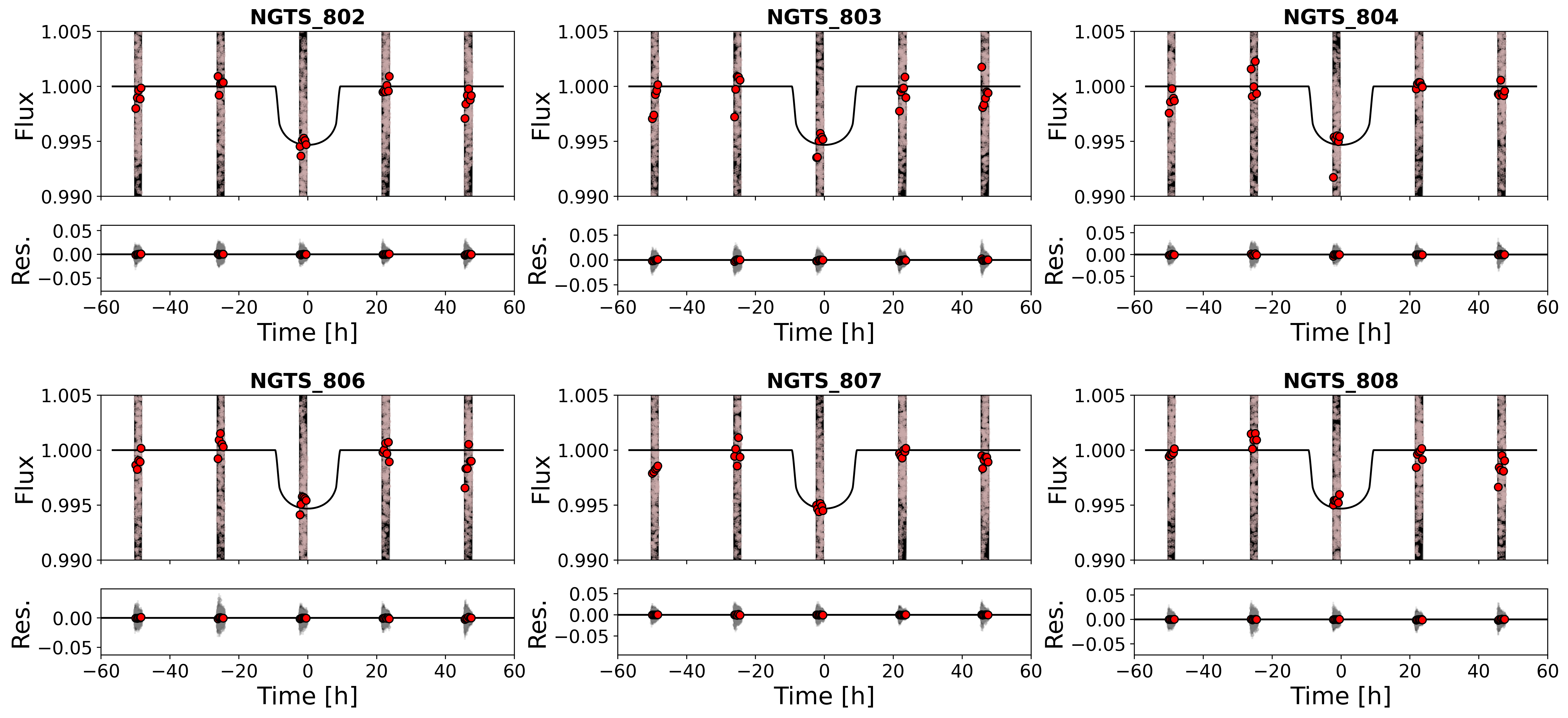}
    \caption{Same as Fig.\ref{fig:lc_neossat} for NGTS light-curves.}
    \label{fig:NGTS_lcs}
\end{figure}

\begin{figure}
    \centering
    \includegraphics[width=0.45\textwidth]{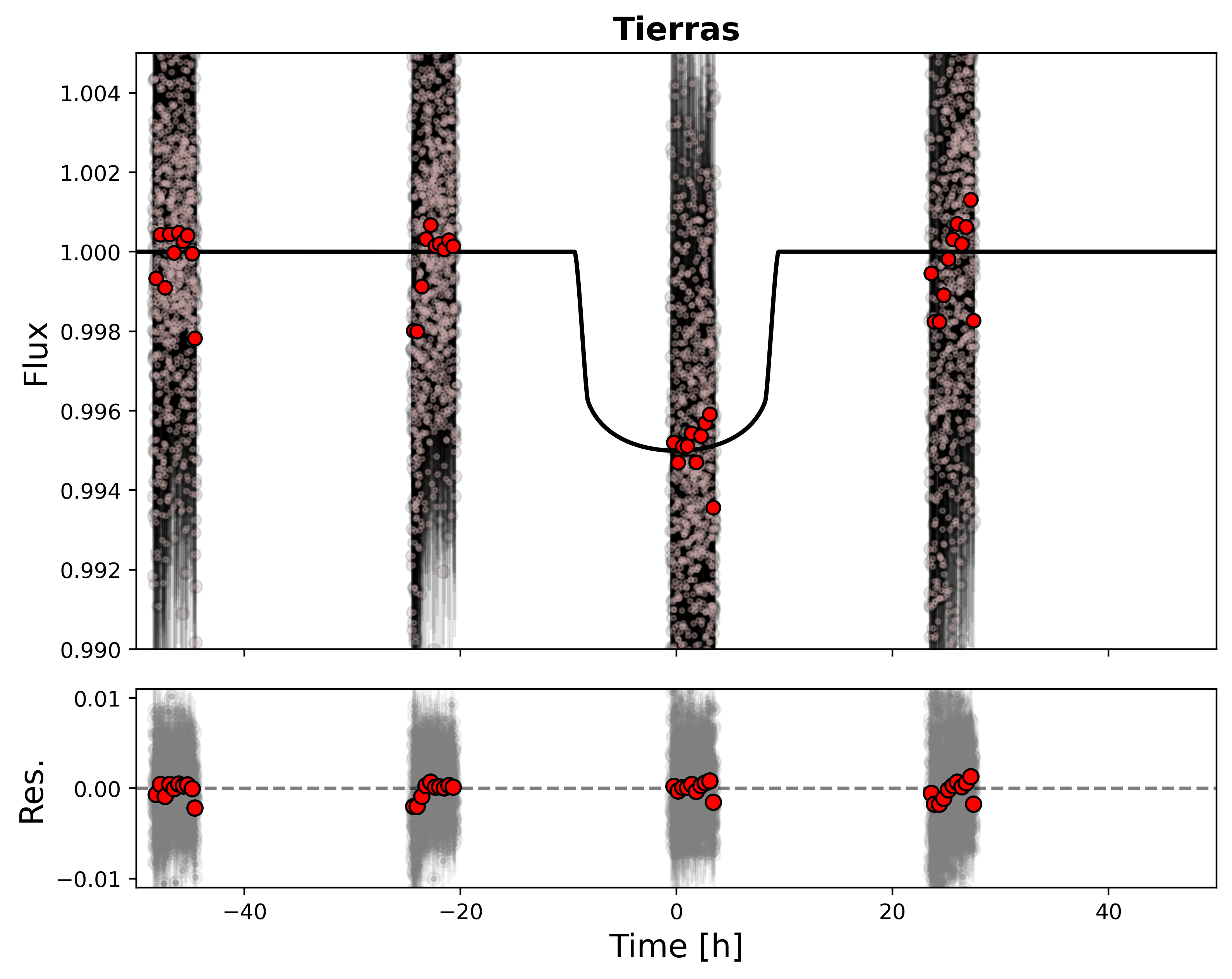}
    \caption{Same as Fig.\ref{fig:lc_neossat} for Tierras light-curve.}
    \label{fig:Tierras_lcs}
\end{figure}

\end{appendix}

\end{document}